\documentclass{llncs}
\makeatletter
\renewcommand*\l@author[2]{}
\renewcommand*\l@title[2]{}
\makeatletter

\usepackage[normalem]{ulem}

\usepackage{framed}
\usepackage{amsmath}
\usepackage{float}
\usepackage{amssymb}
\usepackage{xcolor}
\usepackage{stmaryrd}
\usepackage{wasysym}
\usepackage{pifont}
\usepackage{lscape}
\usepackage{cite}
\usepackage{multicol}
\usepackage{caption}
\usepackage{graphicx}
\usepackage{subfig}
\usepackage{color}
\usepackage{transparent}
\usepackage{import}
\graphicspath{{figures/}}
\usepackage{pbox}

\usepackage[linesnumbered,rightnl,boxed,vlined]{algorithm2e}
\usepackage{comment}

\definecolor{AliceBlue}{rgb}{0.94,0.97,1.00}
\definecolor{AntiqueWhite1}{rgb}{1.00,0.94,0.86}
\definecolor{AntiqueWhite2}{rgb}{0.93,0.87,0.80}
\definecolor{AntiqueWhite3}{rgb}{0.80,0.75,0.69}
\definecolor{AntiqueWhite4}{rgb}{0.55,0.51,0.47}
\definecolor{AntiqueWhite}{rgb}{0.98,0.92,0.84}
\definecolor{BlanchedAlmond}{rgb}{1.00,0.92,0.80}
\definecolor{BlueViolet}{rgb}{0.54,0.17,0.89}
\definecolor{CadetBlue1}{rgb}{0.60,0.96,1.00}
\definecolor{CadetBlue2}{rgb}{0.56,0.90,0.93}
\definecolor{CadetBlue3}{rgb}{0.48,0.77,0.80}
\definecolor{CadetBlue4}{rgb}{0.33,0.53,0.55}
\definecolor{CadetBlue}{rgb}{0.37,0.62,0.63}
\definecolor{CornflowerBlue}{rgb}{0.39,0.58,0.93}
\definecolor{DarkBlue}{rgb}{0.00,0.00,0.55}
\definecolor{DarkCyan}{rgb}{0.00,0.55,0.55}
\definecolor{DarkGoldenrod1}{rgb}{1.00,0.73,0.06}
\definecolor{DarkGoldenrod2}{rgb}{0.93,0.68,0.05}
\definecolor{DarkGoldenrod3}{rgb}{0.80,0.58,0.05}
\definecolor{DarkGoldenrod4}{rgb}{0.55,0.40,0.03}
\definecolor{DarkGoldenrod}{rgb}{0.72,0.53,0.04}
\definecolor{DarkGray}{rgb}{0.66,0.66,0.66}
\definecolor{DarkGreen}{rgb}{0.00,0.39,0.00}
\definecolor{DarkGrey}{rgb}{0.66,0.66,0.66}
\definecolor{DarkKhaki}{rgb}{0.74,0.72,0.42}
\definecolor{DarkMagenta}{rgb}{0.55,0.00,0.55}
\definecolor{DarkOliveGreen1}{rgb}{0.79,1.00,0.44}
\definecolor{DarkOliveGreen2}{rgb}{0.74,0.93,0.41}
\definecolor{DarkOliveGreen3}{rgb}{0.64,0.80,0.35}
\definecolor{DarkOliveGreen4}{rgb}{0.43,0.55,0.24}
\definecolor{DarkOliveGreen}{rgb}{0.33,0.42,0.18}
\definecolor{DarkOrange1}{rgb}{1.00,0.50,0.00}
\definecolor{DarkOrange2}{rgb}{0.93,0.46,0.00}
\definecolor{DarkOrange3}{rgb}{0.80,0.40,0.00}
\definecolor{DarkOrange4}{rgb}{0.55,0.27,0.00}
\definecolor{DarkOrange}{rgb}{1.00,0.55,0.00}
\definecolor{DarkOrchid1}{rgb}{0.75,0.24,1.00}
\definecolor{DarkOrchid2}{rgb}{0.70,0.23,0.93}
\definecolor{DarkOrchid3}{rgb}{0.60,0.20,0.80}
\definecolor{DarkOrchid4}{rgb}{0.41,0.13,0.55}
\definecolor{DarkOrchid}{rgb}{0.60,0.20,0.80}
\definecolor{DarkRed}{rgb}{0.55,0.00,0.00}
\definecolor{DarkSalmon}{rgb}{0.91,0.59,0.48}
\definecolor{DarkSeaGreen1}{rgb}{0.76,1.00,0.76}
\definecolor{DarkSeaGreen2}{rgb}{0.71,0.93,0.71}
\definecolor{DarkSeaGreen3}{rgb}{0.61,0.80,0.61}
\definecolor{DarkSeaGreen4}{rgb}{0.41,0.55,0.41}
\definecolor{DarkSeaGreen}{rgb}{0.56,0.74,0.56}
\definecolor{DarkSlateBlue}{rgb}{0.28,0.24,0.55}
\definecolor{DarkSlateGray1}{rgb}{0.59,1.00,1.00}
\definecolor{DarkSlateGray2}{rgb}{0.55,0.93,0.93}
\definecolor{DarkSlateGray3}{rgb}{0.47,0.80,0.80}
\definecolor{DarkSlateGray4}{rgb}{0.32,0.55,0.55}
\definecolor{DarkSlateGray}{rgb}{0.18,0.31,0.31}
\definecolor{DarkSlateGrey}{rgb}{0.18,0.31,0.31}
\definecolor{DarkTurquoise}{rgb}{0.00,0.81,0.82}
\definecolor{DarkViolet}{rgb}{0.58,0.00,0.83}
\definecolor{DeepPink1}{rgb}{1.00,0.08,0.58}
\definecolor{DeepPink2}{rgb}{0.93,0.07,0.54}
\definecolor{DeepPink3}{rgb}{0.80,0.06,0.46}
\definecolor{DeepPink4}{rgb}{0.55,0.04,0.31}
\definecolor{DeepPink}{rgb}{1.00,0.08,0.58}
\definecolor{DeepSkyBlue1}{rgb}{0.00,0.75,1.00}
\definecolor{DeepSkyBlue2}{rgb}{0.00,0.70,0.93}
\definecolor{DeepSkyBlue3}{rgb}{0.00,0.60,0.80}
\definecolor{DeepSkyBlue4}{rgb}{0.00,0.41,0.55}
\definecolor{DeepSkyBlue}{rgb}{0.00,0.75,1.00}
\definecolor{DimGray}{rgb}{0.41,0.41,0.41}
\definecolor{DimGrey}{rgb}{0.41,0.41,0.41}
\definecolor{DodgerBlue1}{rgb}{0.12,0.56,1.00}
\definecolor{DodgerBlue2}{rgb}{0.11,0.53,0.93}
\definecolor{DodgerBlue3}{rgb}{0.09,0.45,0.80}
\definecolor{DodgerBlue4}{rgb}{0.06,0.31,0.55}
\definecolor{DodgerBlue}{rgb}{0.12,0.56,1.00}
\definecolor{FloralWhite}{rgb}{1.00,0.98,0.94}
\definecolor{ForestGreen}{rgb}{0.13,0.55,0.13}
\definecolor{GhostWhite}{rgb}{0.97,0.97,1.00}
\definecolor{GreenYellow}{rgb}{0.68,1.00,0.18}
\definecolor{HotPink1}{rgb}{1.00,0.43,0.71}
\definecolor{HotPink2}{rgb}{0.93,0.42,0.65}
\definecolor{HotPink3}{rgb}{0.80,0.38,0.56}
\definecolor{HotPink4}{rgb}{0.55,0.23,0.38}
\definecolor{HotPink}{rgb}{1.00,0.41,0.71}
\definecolor{IndianRed1}{rgb}{1.00,0.42,0.42}
\definecolor{IndianRed2}{rgb}{0.93,0.39,0.39}
\definecolor{IndianRed3}{rgb}{0.80,0.33,0.33}
\definecolor{IndianRed4}{rgb}{0.55,0.23,0.23}
\definecolor{IndianRed}{rgb}{0.80,0.36,0.36}
\definecolor{LavenderBlush1}{rgb}{1.00,0.94,0.96}
\definecolor{LavenderBlush2}{rgb}{0.93,0.88,0.90}
\definecolor{LavenderBlush3}{rgb}{0.80,0.76,0.77}
\definecolor{LavenderBlush4}{rgb}{0.55,0.51,0.53}
\definecolor{LavenderBlush}{rgb}{1.00,0.94,0.96}
\definecolor{LawnGreen}{rgb}{0.49,0.99,0.00}
\definecolor{LemonChiffon1}{rgb}{1.00,0.98,0.80}
\definecolor{LemonChiffon2}{rgb}{0.93,0.91,0.75}
\definecolor{LemonChiffon3}{rgb}{0.80,0.79,0.65}
\definecolor{LemonChiffon4}{rgb}{0.55,0.54,0.44}
\definecolor{LemonChiffon}{rgb}{1.00,0.98,0.80}
\definecolor{LightBlue1}{rgb}{0.75,0.94,1.00}
\definecolor{LightBlue2}{rgb}{0.70,0.87,0.93}
\definecolor{LightBlue3}{rgb}{0.60,0.75,0.80}
\definecolor{LightBlue4}{rgb}{0.41,0.51,0.55}
\definecolor{LightBlue}{rgb}{0.68,0.85,0.90}
\definecolor{LightCoral}{rgb}{0.94,0.50,0.50}
\definecolor{LightCyan1}{rgb}{0.88,1.00,1.00}
\definecolor{LightCyan2}{rgb}{0.82,0.93,0.93}
\definecolor{LightCyan3}{rgb}{0.71,0.80,0.80}
\definecolor{LightCyan4}{rgb}{0.48,0.55,0.55}
\definecolor{LightCyan}{rgb}{0.88,1.00,1.00}
\definecolor{LightGoldenrod1}{rgb}{1.00,0.93,0.55}
\definecolor{LightGoldenrod2}{rgb}{0.93,0.86,0.51}
\definecolor{LightGoldenrod3}{rgb}{0.80,0.75,0.44}
\definecolor{LightGoldenrod4}{rgb}{0.55,0.51,0.30}
\definecolor{LightGoldenrodYellow}{rgb}{0.98,0.98,0.82}
\definecolor{LightGoldenrod}{rgb}{0.93,0.87,0.51}
\definecolor{LightGray}{rgb}{0.83,0.83,0.83}
\definecolor{LightGreen}{rgb}{0.56,0.93,0.56}
\definecolor{LightGrey}{rgb}{0.83,0.83,0.83}
\definecolor{LightPink1}{rgb}{1.00,0.68,0.73}
\definecolor{LightPink2}{rgb}{0.93,0.64,0.68}
\definecolor{LightPink3}{rgb}{0.80,0.55,0.58}
\definecolor{LightPink4}{rgb}{0.55,0.37,0.40}
\definecolor{LightPink}{rgb}{1.00,0.71,0.76}
\definecolor{LightSalmon1}{rgb}{1.00,0.63,0.48}
\definecolor{LightSalmon2}{rgb}{0.93,0.58,0.45}
\definecolor{LightSalmon3}{rgb}{0.80,0.51,0.38}
\definecolor{LightSalmon4}{rgb}{0.55,0.34,0.26}
\definecolor{LightSalmon}{rgb}{1.00,0.63,0.48}
\definecolor{LightSeaGreen}{rgb}{0.13,0.70,0.67}
\definecolor{LightSkyBlue1}{rgb}{0.69,0.89,1.00}
\definecolor{LightSkyBlue2}{rgb}{0.64,0.83,0.93}
\definecolor{LightSkyBlue3}{rgb}{0.55,0.71,0.80}
\definecolor{LightSkyBlue4}{rgb}{0.38,0.48,0.55}
\definecolor{LightSkyBlue}{rgb}{0.53,0.81,0.98}
\definecolor{LightSlateBlue}{rgb}{0.52,0.44,1.00}
\definecolor{LightSlateGray}{rgb}{0.47,0.53,0.60}
\definecolor{LightSlateGrey}{rgb}{0.47,0.53,0.60}
\definecolor{LightSteelBlue1}{rgb}{0.79,0.88,1.00}
\definecolor{LightSteelBlue2}{rgb}{0.74,0.82,0.93}
\definecolor{LightSteelBlue3}{rgb}{0.64,0.71,0.80}
\definecolor{LightSteelBlue4}{rgb}{0.43,0.48,0.55}
\definecolor{LightSteelBlue}{rgb}{0.69,0.77,0.87}
\definecolor{LightYellow1}{rgb}{1.00,1.00,0.88}
\definecolor{LightYellow2}{rgb}{0.93,0.93,0.82}
\definecolor{LightYellow3}{rgb}{0.80,0.80,0.71}
\definecolor{LightYellow4}{rgb}{0.55,0.55,0.48}
\definecolor{LightYellow}{rgb}{1.00,1.00,0.88}
\definecolor{LimeGreen}{rgb}{0.20,0.80,0.20}
\definecolor{MediumAquamarine}{rgb}{0.40,0.80,0.67}
\definecolor{MediumBlue}{rgb}{0.00,0.00,0.80}
\definecolor{MediumOrchid1}{rgb}{0.88,0.40,1.00}
\definecolor{MediumOrchid2}{rgb}{0.82,0.37,0.93}
\definecolor{MediumOrchid3}{rgb}{0.71,0.32,0.80}
\definecolor{MediumOrchid4}{rgb}{0.48,0.22,0.55}
\definecolor{MediumOrchid}{rgb}{0.73,0.33,0.83}
\definecolor{MediumPurple1}{rgb}{0.67,0.51,1.00}
\definecolor{MediumPurple2}{rgb}{0.62,0.47,0.93}
\definecolor{MediumPurple3}{rgb}{0.54,0.41,0.80}
\definecolor{MediumPurple4}{rgb}{0.36,0.28,0.55}
\definecolor{MediumPurple}{rgb}{0.58,0.44,0.86}
\definecolor{MediumSeaGreen}{rgb}{0.24,0.70,0.44}
\definecolor{MediumSlateBlue}{rgb}{0.48,0.41,0.93}
\definecolor{MediumSpringGreen}{rgb}{0.00,0.98,0.60}
\definecolor{MediumTurquoise}{rgb}{0.28,0.82,0.80}
\definecolor{MediumVioletRed}{rgb}{0.78,0.08,0.52}
\definecolor{MidnightBlue}{rgb}{0.10,0.10,0.44}
\definecolor{MintCream}{rgb}{0.96,1.00,0.98}
\definecolor{MistyRose1}{rgb}{1.00,0.89,0.88}
\definecolor{MistyRose2}{rgb}{0.93,0.84,0.82}
\definecolor{MistyRose3}{rgb}{0.80,0.72,0.71}
\definecolor{MistyRose4}{rgb}{0.55,0.49,0.48}
\definecolor{MistyRose}{rgb}{1.00,0.89,0.88}
\definecolor{NavajoWhite1}{rgb}{1.00,0.87,0.68}
\definecolor{NavajoWhite2}{rgb}{0.93,0.81,0.63}
\definecolor{NavajoWhite3}{rgb}{0.80,0.70,0.55}
\definecolor{NavajoWhite4}{rgb}{0.55,0.47,0.37}
\definecolor{NavajoWhite}{rgb}{1.00,0.87,0.68}
\definecolor{NavyBlue}{rgb}{0.00,0.00,0.50}
\definecolor{OldLace}{rgb}{0.99,0.96,0.90}
\definecolor{OliveDrab1}{rgb}{0.75,1.00,0.24}
\definecolor{OliveDrab2}{rgb}{0.70,0.93,0.23}
\definecolor{OliveDrab3}{rgb}{0.60,0.80,0.20}
\definecolor{OliveDrab4}{rgb}{0.41,0.55,0.13}
\definecolor{OliveDrab}{rgb}{0.42,0.56,0.14}
\definecolor{OrangeRed1}{rgb}{1.00,0.27,0.00}
\definecolor{OrangeRed2}{rgb}{0.93,0.25,0.00}
\definecolor{OrangeRed3}{rgb}{0.80,0.22,0.00}
\definecolor{OrangeRed4}{rgb}{0.55,0.15,0.00}
\definecolor{OrangeRed}{rgb}{1.00,0.27,0.00}
\definecolor{PaleGoldenrod}{rgb}{0.93,0.91,0.67}
\definecolor{PaleGreen1}{rgb}{0.60,1.00,0.60}
\definecolor{PaleGreen2}{rgb}{0.56,0.93,0.56}
\definecolor{PaleGreen3}{rgb}{0.49,0.80,0.49}
\definecolor{PaleGreen4}{rgb}{0.33,0.55,0.33}
\definecolor{PaleGreen}{rgb}{0.60,0.98,0.60}
\definecolor{PaleTurquoise1}{rgb}{0.73,1.00,1.00}
\definecolor{PaleTurquoise2}{rgb}{0.68,0.93,0.93}
\definecolor{PaleTurquoise3}{rgb}{0.59,0.80,0.80}
\definecolor{PaleTurquoise4}{rgb}{0.40,0.55,0.55}
\definecolor{PaleTurquoise}{rgb}{0.69,0.93,0.93}
\definecolor{PaleVioletRed1}{rgb}{1.00,0.51,0.67}
\definecolor{PaleVioletRed2}{rgb}{0.93,0.47,0.62}
\definecolor{PaleVioletRed3}{rgb}{0.80,0.41,0.54}
\definecolor{PaleVioletRed4}{rgb}{0.55,0.28,0.36}
\definecolor{PaleVioletRed}{rgb}{0.86,0.44,0.58}
\definecolor{PapayaWhip}{rgb}{1.00,0.94,0.84}
\definecolor{PeachPuff1}{rgb}{1.00,0.85,0.73}
\definecolor{PeachPuff2}{rgb}{0.93,0.80,0.68}
\definecolor{PeachPuff3}{rgb}{0.80,0.69,0.58}
\definecolor{PeachPuff4}{rgb}{0.55,0.47,0.40}
\definecolor{PeachPuff}{rgb}{1.00,0.85,0.73}
\definecolor{PowderBlue}{rgb}{0.69,0.88,0.90}
\definecolor{RosyBrown1}{rgb}{1.00,0.76,0.76}
\definecolor{RosyBrown2}{rgb}{0.93,0.71,0.71}
\definecolor{RosyBrown3}{rgb}{0.80,0.61,0.61}
\definecolor{RosyBrown4}{rgb}{0.55,0.41,0.41}
\definecolor{RosyBrown}{rgb}{0.74,0.56,0.56}
\definecolor{RoyalBlue1}{rgb}{0.28,0.46,1.00}
\definecolor{RoyalBlue2}{rgb}{0.26,0.43,0.93}
\definecolor{RoyalBlue3}{rgb}{0.23,0.37,0.80}
\definecolor{RoyalBlue4}{rgb}{0.15,0.25,0.55}
\definecolor{RoyalBlue}{rgb}{0.25,0.41,0.88}
\definecolor{SaddleBrown}{rgb}{0.55,0.27,0.07}
\definecolor{SandyBrown}{rgb}{0.96,0.64,0.38}
\definecolor{SeaGreen1}{rgb}{0.33,1.00,0.62}
\definecolor{SeaGreen2}{rgb}{0.31,0.93,0.58}
\definecolor{SeaGreen3}{rgb}{0.26,0.80,0.50}
\definecolor{SeaGreen4}{rgb}{0.18,0.55,0.34}
\definecolor{SeaGreen}{rgb}{0.18,0.55,0.34}
\definecolor{SkyBlue1}{rgb}{0.53,0.81,1.00}
\definecolor{SkyBlue2}{rgb}{0.49,0.75,0.93}
\definecolor{SkyBlue3}{rgb}{0.42,0.65,0.80}
\definecolor{SkyBlue4}{rgb}{0.29,0.44,0.55}
\definecolor{SkyBlue}{rgb}{0.53,0.81,0.92}
\definecolor{SlateBlue1}{rgb}{0.51,0.44,1.00}
\definecolor{SlateBlue2}{rgb}{0.48,0.40,0.93}
\definecolor{SlateBlue3}{rgb}{0.41,0.35,0.80}
\definecolor{SlateBlue4}{rgb}{0.28,0.24,0.55}
\definecolor{SlateBlue}{rgb}{0.42,0.35,0.80}
\definecolor{SlateGray1}{rgb}{0.78,0.89,1.00}
\definecolor{SlateGray2}{rgb}{0.73,0.83,0.93}
\definecolor{SlateGray3}{rgb}{0.62,0.71,0.80}
\definecolor{SlateGray4}{rgb}{0.42,0.48,0.55}
\definecolor{SlateGray}{rgb}{0.44,0.50,0.56}
\definecolor{SlateGrey}{rgb}{0.44,0.50,0.56}
\definecolor{SpringGreen1}{rgb}{0.00,1.00,0.50}
\definecolor{SpringGreen2}{rgb}{0.00,0.93,0.46}
\definecolor{SpringGreen3}{rgb}{0.00,0.80,0.40}
\definecolor{SpringGreen4}{rgb}{0.00,0.55,0.27}
\definecolor{SpringGreen}{rgb}{0.00,1.00,0.50}
\definecolor{SteelBlue1}{rgb}{0.39,0.72,1.00}
\definecolor{SteelBlue2}{rgb}{0.36,0.67,0.93}
\definecolor{SteelBlue3}{rgb}{0.31,0.58,0.80}
\definecolor{SteelBlue4}{rgb}{0.21,0.39,0.55}
\definecolor{SteelBlue}{rgb}{0.27,0.51,0.71}
\definecolor{VioletRed1}{rgb}{1.00,0.24,0.59}
\definecolor{VioletRed2}{rgb}{0.93,0.23,0.55}
\definecolor{VioletRed3}{rgb}{0.80,0.20,0.47}
\definecolor{VioletRed4}{rgb}{0.55,0.13,0.32}
\definecolor{VioletRed}{rgb}{0.82,0.13,0.56}
\definecolor{WhiteSmoke}{rgb}{0.96,0.96,0.96}
\definecolor{YellowGreen}{rgb}{0.60,0.80,0.20}
\definecolor{aliceblue}{rgb}{0.94,0.97,1.00}
\definecolor{antiquewhite}{rgb}{0.98,0.92,0.84}
\definecolor{aquamarine1}{rgb}{0.50,1.00,0.83}
\definecolor{aquamarine2}{rgb}{0.46,0.93,0.78}
\definecolor{aquamarine3}{rgb}{0.40,0.80,0.67}
\definecolor{aquamarine4}{rgb}{0.27,0.55,0.45}
\definecolor{aquamarine}{rgb}{0.50,1.00,0.83}
\definecolor{azure1}{rgb}{0.94,1.00,1.00}
\definecolor{azure2}{rgb}{0.88,0.93,0.93}
\definecolor{azure3}{rgb}{0.76,0.80,0.80}
\definecolor{azure4}{rgb}{0.51,0.55,0.55}
\definecolor{azure}{rgb}{0.94,1.00,1.00}
\definecolor{beige}{rgb}{0.96,0.96,0.86}
\definecolor{bisque1}{rgb}{1.00,0.89,0.77}
\definecolor{bisque2}{rgb}{0.93,0.84,0.72}
\definecolor{bisque3}{rgb}{0.80,0.72,0.62}
\definecolor{bisque4}{rgb}{0.55,0.49,0.42}
\definecolor{bisque}{rgb}{1.00,0.89,0.77}
\definecolor{black}{rgb}{0.00,0.00,0.00}
\definecolor{blanchedalmond}{rgb}{1.00,0.92,0.80}
\definecolor{blue1}{rgb}{0.00,0.00,1.00}
\definecolor{blue2}{rgb}{0.00,0.00,0.93}
\definecolor{blue3}{rgb}{0.00,0.00,0.80}
\definecolor{blue4}{rgb}{0.00,0.00,0.55}
\definecolor{blueviolet}{rgb}{0.54,0.17,0.89}
\definecolor{blue}{rgb}{0.00,0.00,1.00}
\definecolor{brown1}{rgb}{1.00,0.25,0.25}
\definecolor{brown2}{rgb}{0.93,0.23,0.23}
\definecolor{brown3}{rgb}{0.80,0.20,0.20}
\definecolor{brown4}{rgb}{0.55,0.14,0.14}
\definecolor{brown}{rgb}{0.65,0.16,0.16}
\definecolor{burlywood1}{rgb}{1.00,0.83,0.61}
\definecolor{burlywood2}{rgb}{0.93,0.77,0.57}
\definecolor{burlywood3}{rgb}{0.80,0.67,0.49}
\definecolor{burlywood4}{rgb}{0.55,0.45,0.33}
\definecolor{burlywood}{rgb}{0.87,0.72,0.53}
\definecolor{cadetblue}{rgb}{0.37,0.62,0.63}
\definecolor{chartreuse1}{rgb}{0.50,1.00,0.00}
\definecolor{chartreuse2}{rgb}{0.46,0.93,0.00}
\definecolor{chartreuse3}{rgb}{0.40,0.80,0.00}
\definecolor{chartreuse4}{rgb}{0.27,0.55,0.00}
\definecolor{chartreuse}{rgb}{0.50,1.00,0.00}
\definecolor{chocolate1}{rgb}{1.00,0.50,0.14}
\definecolor{chocolate2}{rgb}{0.93,0.46,0.13}
\definecolor{chocolate3}{rgb}{0.80,0.40,0.11}
\definecolor{chocolate4}{rgb}{0.55,0.27,0.07}
\definecolor{chocolate}{rgb}{0.82,0.41,0.12}
\definecolor{coral1}{rgb}{1.00,0.45,0.34}
\definecolor{coral2}{rgb}{0.93,0.42,0.31}
\definecolor{coral3}{rgb}{0.80,0.36,0.27}
\definecolor{coral4}{rgb}{0.55,0.24,0.18}
\definecolor{coral}{rgb}{1.00,0.50,0.31}
\definecolor{cornflowerblue}{rgb}{0.39,0.58,0.93}
\definecolor{cornsilk1}{rgb}{1.00,0.97,0.86}
\definecolor{cornsilk2}{rgb}{0.93,0.91,0.80}
\definecolor{cornsilk3}{rgb}{0.80,0.78,0.69}
\definecolor{cornsilk4}{rgb}{0.55,0.53,0.47}
\definecolor{cornsilk}{rgb}{1.00,0.97,0.86}
\definecolor{cyan1}{rgb}{0.00,1.00,1.00}
\definecolor{cyan2}{rgb}{0.00,0.93,0.93}
\definecolor{cyan3}{rgb}{0.00,0.80,0.80}
\definecolor{cyan4}{rgb}{0.00,0.55,0.55}
\definecolor{cyan}{rgb}{0.00,1.00,1.00}
\definecolor{darkblue}{rgb}{0.00,0.00,0.55}
\definecolor{darkcyan}{rgb}{0.00,0.55,0.55}
\definecolor{darkgoldenrod}{rgb}{0.72,0.53,0.04}
\definecolor{darkgray}{rgb}{0.66,0.66,0.66}
\definecolor{darkgreen}{rgb}{0.00,0.39,0.00}
\definecolor{darkgrey}{rgb}{0.66,0.66,0.66}
\definecolor{darkkhaki}{rgb}{0.74,0.72,0.42}
\definecolor{darkmagenta}{rgb}{0.55,0.00,0.55}
\definecolor{darkolive}{rgb}{0.33,0.42,0.18}
\definecolor{darkorange}{rgb}{1.00,0.55,0.00}
\definecolor{darkorchid}{rgb}{0.60,0.20,0.80}
\definecolor{darkred}{rgb}{0.55,0.00,0.00}
\definecolor{darksalmon}{rgb}{0.91,0.59,0.48}
\definecolor{darksea}{rgb}{0.56,0.74,0.56}
\definecolor{darkslate}{rgb}{0.18,0.31,0.31}
\definecolor{darkslate}{rgb}{0.18,0.31,0.31}
\definecolor{darkslate}{rgb}{0.28,0.24,0.55}
\definecolor{darkturquoise}{rgb}{0.00,0.81,0.82}
\definecolor{darkviolet}{rgb}{0.58,0.00,0.83}
\definecolor{deeppink}{rgb}{1.00,0.08,0.58}
\definecolor{deepsky}{rgb}{0.00,0.75,1.00}
\definecolor{dimgray}{rgb}{0.41,0.41,0.41}
\definecolor{dimgrey}{rgb}{0.41,0.41,0.41}
\definecolor{dodgerblue}{rgb}{0.12,0.56,1.00}
\definecolor{firebrick1}{rgb}{1.00,0.19,0.19}
\definecolor{firebrick2}{rgb}{0.93,0.17,0.17}
\definecolor{firebrick3}{rgb}{0.80,0.15,0.15}
\definecolor{firebrick4}{rgb}{0.55,0.10,0.10}
\definecolor{firebrick}{rgb}{0.70,0.13,0.13}
\definecolor{floralwhite}{rgb}{1.00,0.98,0.94}
\definecolor{forestgreen}{rgb}{0.13,0.55,0.13}
\definecolor{gainsboro}{rgb}{0.86,0.86,0.86}
\definecolor{ghostwhite}{rgb}{0.97,0.97,1.00}
\definecolor{gold1}{rgb}{1.00,0.84,0.00}
\definecolor{gold2}{rgb}{0.93,0.79,0.00}
\definecolor{gold3}{rgb}{0.80,0.68,0.00}
\definecolor{gold4}{rgb}{0.55,0.46,0.00}
\definecolor{goldenrod1}{rgb}{1.00,0.76,0.15}
\definecolor{goldenrod2}{rgb}{0.93,0.71,0.13}
\definecolor{goldenrod3}{rgb}{0.80,0.61,0.11}
\definecolor{goldenrod4}{rgb}{0.55,0.41,0.08}
\definecolor{goldenrod}{rgb}{0.85,0.65,0.13}
\definecolor{gold}{rgb}{1.00,0.84,0.00}
\definecolor{gray0}{rgb}{0.00,0.00,0.00}
\definecolor{gray100}{rgb}{1.00,1.00,1.00}
\definecolor{gray10}{rgb}{0.10,0.10,0.10}
\definecolor{gray11}{rgb}{0.11,0.11,0.11}
\definecolor{gray12}{rgb}{0.12,0.12,0.12}
\definecolor{gray13}{rgb}{0.13,0.13,0.13}
\definecolor{gray14}{rgb}{0.14,0.14,0.14}
\definecolor{gray15}{rgb}{0.15,0.15,0.15}
\definecolor{gray16}{rgb}{0.16,0.16,0.16}
\definecolor{gray17}{rgb}{0.17,0.17,0.17}
\definecolor{gray18}{rgb}{0.18,0.18,0.18}
\definecolor{gray19}{rgb}{0.19,0.19,0.19}
\definecolor{gray1}{rgb}{0.01,0.01,0.01}
\definecolor{gray20}{rgb}{0.20,0.20,0.20}
\definecolor{gray21}{rgb}{0.21,0.21,0.21}
\definecolor{gray22}{rgb}{0.22,0.22,0.22}
\definecolor{gray23}{rgb}{0.23,0.23,0.23}
\definecolor{gray24}{rgb}{0.24,0.24,0.24}
\definecolor{gray25}{rgb}{0.25,0.25,0.25}
\definecolor{gray26}{rgb}{0.26,0.26,0.26}
\definecolor{gray27}{rgb}{0.27,0.27,0.27}
\definecolor{gray28}{rgb}{0.28,0.28,0.28}
\definecolor{gray29}{rgb}{0.29,0.29,0.29}
\definecolor{gray2}{rgb}{0.02,0.02,0.02}
\definecolor{gray30}{rgb}{0.30,0.30,0.30}
\definecolor{gray31}{rgb}{0.31,0.31,0.31}
\definecolor{gray32}{rgb}{0.32,0.32,0.32}
\definecolor{gray33}{rgb}{0.33,0.33,0.33}
\definecolor{gray34}{rgb}{0.34,0.34,0.34}
\definecolor{gray35}{rgb}{0.35,0.35,0.35}
\definecolor{gray36}{rgb}{0.36,0.36,0.36}
\definecolor{gray37}{rgb}{0.37,0.37,0.37}
\definecolor{gray38}{rgb}{0.38,0.38,0.38}
\definecolor{gray39}{rgb}{0.39,0.39,0.39}
\definecolor{gray3}{rgb}{0.03,0.03,0.03}
\definecolor{gray40}{rgb}{0.40,0.40,0.40}
\definecolor{gray41}{rgb}{0.41,0.41,0.41}
\definecolor{gray42}{rgb}{0.42,0.42,0.42}
\definecolor{gray43}{rgb}{0.43,0.43,0.43}
\definecolor{gray44}{rgb}{0.44,0.44,0.44}
\definecolor{gray45}{rgb}{0.45,0.45,0.45}
\definecolor{gray46}{rgb}{0.46,0.46,0.46}
\definecolor{gray47}{rgb}{0.47,0.47,0.47}
\definecolor{gray48}{rgb}{0.48,0.48,0.48}
\definecolor{gray49}{rgb}{0.49,0.49,0.49}
\definecolor{gray4}{rgb}{0.04,0.04,0.04}
\definecolor{gray50}{rgb}{0.50,0.50,0.50}
\definecolor{gray51}{rgb}{0.51,0.51,0.51}
\definecolor{gray52}{rgb}{0.52,0.52,0.52}
\definecolor{gray53}{rgb}{0.53,0.53,0.53}
\definecolor{gray54}{rgb}{0.54,0.54,0.54}
\definecolor{gray55}{rgb}{0.55,0.55,0.55}
\definecolor{gray56}{rgb}{0.56,0.56,0.56}
\definecolor{gray57}{rgb}{0.57,0.57,0.57}
\definecolor{gray58}{rgb}{0.58,0.58,0.58}
\definecolor{gray59}{rgb}{0.59,0.59,0.59}
\definecolor{gray5}{rgb}{0.05,0.05,0.05}
\definecolor{gray60}{rgb}{0.60,0.60,0.60}
\definecolor{gray61}{rgb}{0.61,0.61,0.61}
\definecolor{gray62}{rgb}{0.62,0.62,0.62}
\definecolor{gray63}{rgb}{0.63,0.63,0.63}
\definecolor{gray64}{rgb}{0.64,0.64,0.64}
\definecolor{gray65}{rgb}{0.65,0.65,0.65}
\definecolor{gray66}{rgb}{0.66,0.66,0.66}
\definecolor{gray67}{rgb}{0.67,0.67,0.67}
\definecolor{gray68}{rgb}{0.68,0.68,0.68}
\definecolor{gray69}{rgb}{0.69,0.69,0.69}
\definecolor{gray6}{rgb}{0.06,0.06,0.06}
\definecolor{gray70}{rgb}{0.70,0.70,0.70}
\definecolor{gray71}{rgb}{0.71,0.71,0.71}
\definecolor{gray72}{rgb}{0.72,0.72,0.72}
\definecolor{gray73}{rgb}{0.73,0.73,0.73}
\definecolor{gray74}{rgb}{0.74,0.74,0.74}
\definecolor{gray75}{rgb}{0.75,0.75,0.75}
\definecolor{gray76}{rgb}{0.76,0.76,0.76}
\definecolor{gray77}{rgb}{0.77,0.77,0.77}
\definecolor{gray78}{rgb}{0.78,0.78,0.78}
\definecolor{gray79}{rgb}{0.79,0.79,0.79}
\definecolor{gray7}{rgb}{0.07,0.07,0.07}
\definecolor{gray80}{rgb}{0.80,0.80,0.80}
\definecolor{gray81}{rgb}{0.81,0.81,0.81}
\definecolor{gray82}{rgb}{0.82,0.82,0.82}
\definecolor{gray83}{rgb}{0.83,0.83,0.83}
\definecolor{gray84}{rgb}{0.84,0.84,0.84}
\definecolor{gray85}{rgb}{0.85,0.85,0.85}
\definecolor{gray86}{rgb}{0.86,0.86,0.86}
\definecolor{gray87}{rgb}{0.87,0.87,0.87}
\definecolor{gray88}{rgb}{0.88,0.88,0.88}
\definecolor{gray89}{rgb}{0.89,0.89,0.89}
\definecolor{gray8}{rgb}{0.08,0.08,0.08}
\definecolor{gray90}{rgb}{0.90,0.90,0.90}
\definecolor{gray91}{rgb}{0.91,0.91,0.91}
\definecolor{gray92}{rgb}{0.92,0.92,0.92}
\definecolor{gray93}{rgb}{0.93,0.93,0.93}
\definecolor{gray94}{rgb}{0.94,0.94,0.94}
\definecolor{gray95}{rgb}{0.95,0.95,0.95}
\definecolor{gray96}{rgb}{0.96,0.96,0.96}
\definecolor{gray97}{rgb}{0.97,0.97,0.97}
\definecolor{gray98}{rgb}{0.98,0.98,0.98}
\definecolor{gray99}{rgb}{0.99,0.99,0.99}
\definecolor{gray9}{rgb}{0.09,0.09,0.09}
\definecolor{gray}{rgb}{0.75,0.75,0.75}
\definecolor{green1}{rgb}{0.00,1.00,0.00}
\definecolor{green2}{rgb}{0.00,0.93,0.00}
\definecolor{green3}{rgb}{0.00,0.80,0.00}
\definecolor{green4}{rgb}{0.00,0.55,0.00}
\definecolor{greenyellow}{rgb}{0.68,1.00,0.18}
\definecolor{green}{rgb}{0.00,1.00,0.00}
\definecolor{grey0}{rgb}{0.00,0.00,0.00}
\definecolor{grey100}{rgb}{1.00,1.00,1.00}
\definecolor{grey10}{rgb}{0.10,0.10,0.10}
\definecolor{grey11}{rgb}{0.11,0.11,0.11}
\definecolor{grey12}{rgb}{0.12,0.12,0.12}
\definecolor{grey13}{rgb}{0.13,0.13,0.13}
\definecolor{grey14}{rgb}{0.14,0.14,0.14}
\definecolor{grey15}{rgb}{0.15,0.15,0.15}
\definecolor{grey16}{rgb}{0.16,0.16,0.16}
\definecolor{grey17}{rgb}{0.17,0.17,0.17}
\definecolor{grey18}{rgb}{0.18,0.18,0.18}
\definecolor{grey19}{rgb}{0.19,0.19,0.19}
\definecolor{grey1}{rgb}{0.01,0.01,0.01}
\definecolor{grey20}{rgb}{0.20,0.20,0.20}
\definecolor{grey21}{rgb}{0.21,0.21,0.21}
\definecolor{grey22}{rgb}{0.22,0.22,0.22}
\definecolor{grey23}{rgb}{0.23,0.23,0.23}
\definecolor{grey24}{rgb}{0.24,0.24,0.24}
\definecolor{grey25}{rgb}{0.25,0.25,0.25}
\definecolor{grey26}{rgb}{0.26,0.26,0.26}
\definecolor{grey27}{rgb}{0.27,0.27,0.27}
\definecolor{grey28}{rgb}{0.28,0.28,0.28}
\definecolor{grey29}{rgb}{0.29,0.29,0.29}
\definecolor{grey2}{rgb}{0.02,0.02,0.02}
\definecolor{grey30}{rgb}{0.30,0.30,0.30}
\definecolor{grey31}{rgb}{0.31,0.31,0.31}
\definecolor{grey32}{rgb}{0.32,0.32,0.32}
\definecolor{grey33}{rgb}{0.33,0.33,0.33}
\definecolor{grey34}{rgb}{0.34,0.34,0.34}
\definecolor{grey35}{rgb}{0.35,0.35,0.35}
\definecolor{grey36}{rgb}{0.36,0.36,0.36}
\definecolor{grey37}{rgb}{0.37,0.37,0.37}
\definecolor{grey38}{rgb}{0.38,0.38,0.38}
\definecolor{grey39}{rgb}{0.39,0.39,0.39}
\definecolor{grey3}{rgb}{0.03,0.03,0.03}
\definecolor{grey40}{rgb}{0.40,0.40,0.40}
\definecolor{grey41}{rgb}{0.41,0.41,0.41}
\definecolor{grey42}{rgb}{0.42,0.42,0.42}
\definecolor{grey43}{rgb}{0.43,0.43,0.43}
\definecolor{grey44}{rgb}{0.44,0.44,0.44}
\definecolor{grey45}{rgb}{0.45,0.45,0.45}
\definecolor{grey46}{rgb}{0.46,0.46,0.46}
\definecolor{grey47}{rgb}{0.47,0.47,0.47}
\definecolor{grey48}{rgb}{0.48,0.48,0.48}
\definecolor{grey49}{rgb}{0.49,0.49,0.49}
\definecolor{grey4}{rgb}{0.04,0.04,0.04}
\definecolor{grey50}{rgb}{0.50,0.50,0.50}
\definecolor{grey51}{rgb}{0.51,0.51,0.51}
\definecolor{grey52}{rgb}{0.52,0.52,0.52}
\definecolor{grey53}{rgb}{0.53,0.53,0.53}
\definecolor{grey54}{rgb}{0.54,0.54,0.54}
\definecolor{grey55}{rgb}{0.55,0.55,0.55}
\definecolor{grey56}{rgb}{0.56,0.56,0.56}
\definecolor{grey57}{rgb}{0.57,0.57,0.57}
\definecolor{grey58}{rgb}{0.58,0.58,0.58}
\definecolor{grey59}{rgb}{0.59,0.59,0.59}
\definecolor{grey5}{rgb}{0.05,0.05,0.05}
\definecolor{grey60}{rgb}{0.60,0.60,0.60}
\definecolor{grey61}{rgb}{0.61,0.61,0.61}
\definecolor{grey62}{rgb}{0.62,0.62,0.62}
\definecolor{grey63}{rgb}{0.63,0.63,0.63}
\definecolor{grey64}{rgb}{0.64,0.64,0.64}
\definecolor{grey65}{rgb}{0.65,0.65,0.65}
\definecolor{grey66}{rgb}{0.66,0.66,0.66}
\definecolor{grey67}{rgb}{0.67,0.67,0.67}
\definecolor{grey68}{rgb}{0.68,0.68,0.68}
\definecolor{grey69}{rgb}{0.69,0.69,0.69}
\definecolor{grey6}{rgb}{0.06,0.06,0.06}
\definecolor{grey70}{rgb}{0.70,0.70,0.70}
\definecolor{grey71}{rgb}{0.71,0.71,0.71}
\definecolor{grey72}{rgb}{0.72,0.72,0.72}
\definecolor{grey73}{rgb}{0.73,0.73,0.73}
\definecolor{grey74}{rgb}{0.74,0.74,0.74}
\definecolor{grey75}{rgb}{0.75,0.75,0.75}
\definecolor{grey76}{rgb}{0.76,0.76,0.76}
\definecolor{grey77}{rgb}{0.77,0.77,0.77}
\definecolor{grey78}{rgb}{0.78,0.78,0.78}
\definecolor{grey79}{rgb}{0.79,0.79,0.79}
\definecolor{grey7}{rgb}{0.07,0.07,0.07}
\definecolor{grey80}{rgb}{0.80,0.80,0.80}
\definecolor{grey81}{rgb}{0.81,0.81,0.81}
\definecolor{grey82}{rgb}{0.82,0.82,0.82}
\definecolor{grey83}{rgb}{0.83,0.83,0.83}
\definecolor{grey84}{rgb}{0.84,0.84,0.84}
\definecolor{grey85}{rgb}{0.85,0.85,0.85}
\definecolor{grey86}{rgb}{0.86,0.86,0.86}
\definecolor{grey87}{rgb}{0.87,0.87,0.87}
\definecolor{grey88}{rgb}{0.88,0.88,0.88}
\definecolor{grey89}{rgb}{0.89,0.89,0.89}
\definecolor{grey8}{rgb}{0.08,0.08,0.08}
\definecolor{grey90}{rgb}{0.90,0.90,0.90}
\definecolor{grey91}{rgb}{0.91,0.91,0.91}
\definecolor{grey92}{rgb}{0.92,0.92,0.92}
\definecolor{grey93}{rgb}{0.93,0.93,0.93}
\definecolor{grey94}{rgb}{0.94,0.94,0.94}
\definecolor{grey95}{rgb}{0.95,0.95,0.95}
\definecolor{grey96}{rgb}{0.96,0.96,0.96}
\definecolor{grey97}{rgb}{0.97,0.97,0.97}
\definecolor{grey98}{rgb}{0.98,0.98,0.98}
\definecolor{grey99}{rgb}{0.99,0.99,0.99}
\definecolor{grey9}{rgb}{0.09,0.09,0.09}
\definecolor{grey}{rgb}{0.75,0.75,0.75}
\definecolor{honeydew1}{rgb}{0.94,1.00,0.94}
\definecolor{honeydew2}{rgb}{0.88,0.93,0.88}
\definecolor{honeydew3}{rgb}{0.76,0.80,0.76}
\definecolor{honeydew4}{rgb}{0.51,0.55,0.51}
\definecolor{honeydew}{rgb}{0.94,1.00,0.94}
\definecolor{hotpink}{rgb}{1.00,0.41,0.71}
\definecolor{indianred}{rgb}{0.80,0.36,0.36}
\definecolor{ivory1}{rgb}{1.00,1.00,0.94}
\definecolor{ivory2}{rgb}{0.93,0.93,0.88}
\definecolor{ivory3}{rgb}{0.80,0.80,0.76}
\definecolor{ivory4}{rgb}{0.55,0.55,0.51}
\definecolor{ivory}{rgb}{1.00,1.00,0.94}
\definecolor{khaki1}{rgb}{1.00,0.96,0.56}
\definecolor{khaki2}{rgb}{0.93,0.90,0.52}
\definecolor{khaki3}{rgb}{0.80,0.78,0.45}
\definecolor{khaki4}{rgb}{0.55,0.53,0.31}
\definecolor{khaki}{rgb}{0.94,0.90,0.55}
\definecolor{lavenderblush}{rgb}{1.00,0.94,0.96}
\definecolor{lavender}{rgb}{0.90,0.90,0.98}
\definecolor{lawngreen}{rgb}{0.49,0.99,0.00}
\definecolor{lemonchiffon}{rgb}{1.00,0.98,0.80}
\definecolor{lightblue}{rgb}{0.68,0.85,0.90}
\definecolor{lightcoral}{rgb}{0.94,0.50,0.50}
\definecolor{lightcyan}{rgb}{0.88,1.00,1.00}
\definecolor{lightgoldenrod}{rgb}{0.93,0.87,0.51}
\definecolor{lightgoldenrod}{rgb}{0.98,0.98,0.82}
\definecolor{lightgray}{rgb}{0.83,0.83,0.83}
\definecolor{lightgreen}{rgb}{0.56,0.93,0.56}
\definecolor{lightgrey}{rgb}{0.83,0.83,0.83}
\definecolor{lightpink}{rgb}{1.00,0.71,0.76}
\definecolor{lightsalmon}{rgb}{1.00,0.63,0.48}
\definecolor{lightsea}{rgb}{0.13,0.70,0.67}
\definecolor{lightsky}{rgb}{0.53,0.81,0.98}
\definecolor{lightslate}{rgb}{0.47,0.53,0.60}
\definecolor{lightslate}{rgb}{0.47,0.53,0.60}
\definecolor{lightslate}{rgb}{0.52,0.44,1.00}
\definecolor{lightsteel}{rgb}{0.69,0.77,0.87}
\definecolor{lightyellow}{rgb}{1.00,1.00,0.88}
\definecolor{limegreen}{rgb}{0.20,0.80,0.20}
\definecolor{linen}{rgb}{0.98,0.94,0.90}
\definecolor{magenta1}{rgb}{1.00,0.00,1.00}
\definecolor{magenta2}{rgb}{0.93,0.00,0.93}
\definecolor{magenta3}{rgb}{0.80,0.00,0.80}
\definecolor{magenta4}{rgb}{0.55,0.00,0.55}
\definecolor{magenta}{rgb}{1.00,0.00,1.00}
\definecolor{maroon1}{rgb}{1.00,0.20,0.70}
\definecolor{maroon2}{rgb}{0.93,0.19,0.65}
\definecolor{maroon3}{rgb}{0.80,0.16,0.56}
\definecolor{maroon4}{rgb}{0.55,0.11,0.38}
\definecolor{maroon}{rgb}{0.69,0.19,0.38}
\definecolor{mediumaquamarine}{rgb}{0.40,0.80,0.67}
\definecolor{mediumblue}{rgb}{0.00,0.00,0.80}
\definecolor{mediumorchid}{rgb}{0.73,0.33,0.83}
\definecolor{mediumpurple}{rgb}{0.58,0.44,0.86}
\definecolor{mediumsea}{rgb}{0.24,0.70,0.44}
\definecolor{mediumslate}{rgb}{0.48,0.41,0.93}
\definecolor{mediumspring}{rgb}{0.00,0.98,0.60}
\definecolor{mediumturquoise}{rgb}{0.28,0.82,0.80}
\definecolor{mediumviolet}{rgb}{0.78,0.08,0.52}
\definecolor{midnightblue}{rgb}{0.10,0.10,0.44}
\definecolor{mintcream}{rgb}{0.96,1.00,0.98}
\definecolor{mistyrose}{rgb}{1.00,0.89,0.88}
\definecolor{moccasin}{rgb}{1.00,0.89,0.71}
\definecolor{navajowhite}{rgb}{1.00,0.87,0.68}
\definecolor{navyblue}{rgb}{0.00,0.00,0.50}
\definecolor{navy}{rgb}{0.00,0.00,0.50}
\definecolor{oldlace}{rgb}{0.99,0.96,0.90}
\definecolor{olivedrab}{rgb}{0.42,0.56,0.14}
\definecolor{orange1}{rgb}{1.00,0.65,0.00}
\definecolor{orange2}{rgb}{0.93,0.60,0.00}
\definecolor{orange3}{rgb}{0.80,0.52,0.00}
\definecolor{orange4}{rgb}{0.55,0.35,0.00}
\definecolor{orangered}{rgb}{1.00,0.27,0.00}
\definecolor{orange}{rgb}{1.00,0.65,0.00}
\definecolor{orchid1}{rgb}{1.00,0.51,0.98}
\definecolor{orchid2}{rgb}{0.93,0.48,0.91}
\definecolor{orchid3}{rgb}{0.80,0.41,0.79}
\definecolor{orchid4}{rgb}{0.55,0.28,0.54}
\definecolor{orchid}{rgb}{0.85,0.44,0.84}
\definecolor{palegoldenrod}{rgb}{0.93,0.91,0.67}
\definecolor{palegreen}{rgb}{0.60,0.98,0.60}
\definecolor{paleturquoise}{rgb}{0.69,0.93,0.93}
\definecolor{paleviolet}{rgb}{0.86,0.44,0.58}
\definecolor{papayawhip}{rgb}{1.00,0.94,0.84}
\definecolor{peachpuff}{rgb}{1.00,0.85,0.73}
\definecolor{peru}{rgb}{0.80,0.52,0.25}
\definecolor{pink1}{rgb}{1.00,0.71,0.77}
\definecolor{pink2}{rgb}{0.93,0.66,0.72}
\definecolor{pink3}{rgb}{0.80,0.57,0.62}
\definecolor{pink4}{rgb}{0.55,0.39,0.42}
\definecolor{pink}{rgb}{1.00,0.75,0.80}
\definecolor{plum1}{rgb}{1.00,0.73,1.00}
\definecolor{plum2}{rgb}{0.93,0.68,0.93}
\definecolor{plum3}{rgb}{0.80,0.59,0.80}
\definecolor{plum4}{rgb}{0.55,0.40,0.55}
\definecolor{plum}{rgb}{0.87,0.63,0.87}
\definecolor{powderblue}{rgb}{0.69,0.88,0.90}
\definecolor{purple1}{rgb}{0.61,0.19,1.00}
\definecolor{purple2}{rgb}{0.57,0.17,0.93}
\definecolor{purple3}{rgb}{0.49,0.15,0.80}
\definecolor{purple4}{rgb}{0.33,0.10,0.55}
\definecolor{purple}{rgb}{0.63,0.13,0.94}
\definecolor{red1}{rgb}{1.00,0.00,0.00}
\definecolor{red2}{rgb}{0.93,0.00,0.00}
\definecolor{red3}{rgb}{0.80,0.00,0.00}
\definecolor{red4}{rgb}{0.55,0.00,0.00}
\definecolor{red}{rgb}{1.00,0.00,0.00}
\definecolor{rosybrown}{rgb}{0.74,0.56,0.56}
\definecolor{royalblue}{rgb}{0.25,0.41,0.88}
\definecolor{saddlebrown}{rgb}{0.55,0.27,0.07}
\definecolor{salmon1}{rgb}{1.00,0.55,0.41}
\definecolor{salmon2}{rgb}{0.93,0.51,0.38}
\definecolor{salmon3}{rgb}{0.80,0.44,0.33}
\definecolor{salmon4}{rgb}{0.55,0.30,0.22}
\definecolor{salmon}{rgb}{0.98,0.50,0.45}
\definecolor{sandybrown}{rgb}{0.96,0.64,0.38}
\definecolor{seagreen}{rgb}{0.18,0.55,0.34}
\definecolor{seashell1}{rgb}{1.00,0.96,0.93}
\definecolor{seashell2}{rgb}{0.93,0.90,0.87}
\definecolor{seashell3}{rgb}{0.80,0.77,0.75}
\definecolor{seashell4}{rgb}{0.55,0.53,0.51}
\definecolor{seashell}{rgb}{1.00,0.96,0.93}
\definecolor{sienna1}{rgb}{1.00,0.51,0.28}
\definecolor{sienna2}{rgb}{0.93,0.47,0.26}
\definecolor{sienna3}{rgb}{0.80,0.41,0.22}
\definecolor{sienna4}{rgb}{0.55,0.28,0.15}
\definecolor{sienna}{rgb}{0.63,0.32,0.18}
\definecolor{skyblue}{rgb}{0.53,0.81,0.92}
\definecolor{slateblue}{rgb}{0.42,0.35,0.80}
\definecolor{slategray}{rgb}{0.44,0.50,0.56}
\definecolor{slategrey}{rgb}{0.44,0.50,0.56}
\definecolor{snow1}{rgb}{1.00,0.98,0.98}
\definecolor{snow2}{rgb}{0.93,0.91,0.91}
\definecolor{snow3}{rgb}{0.80,0.79,0.79}
\definecolor{snow4}{rgb}{0.55,0.54,0.54}
\definecolor{snow}{rgb}{1.00,0.98,0.98}
\definecolor{springgreen}{rgb}{0.00,1.00,0.50}
\definecolor{steelblue}{rgb}{0.27,0.51,0.71}
\definecolor{tan1}{rgb}{1.00,0.65,0.31}
\definecolor{tan2}{rgb}{0.93,0.60,0.29}
\definecolor{tan3}{rgb}{0.80,0.52,0.25}
\definecolor{tan4}{rgb}{0.55,0.35,0.17}
\definecolor{tan}{rgb}{0.82,0.71,0.55}
\definecolor{thistle1}{rgb}{1.00,0.88,1.00}
\definecolor{thistle2}{rgb}{0.93,0.82,0.93}
\definecolor{thistle3}{rgb}{0.80,0.71,0.80}
\definecolor{thistle4}{rgb}{0.55,0.48,0.55}
\definecolor{thistle}{rgb}{0.85,0.75,0.85}
\definecolor{tomato1}{rgb}{1.00,0.39,0.28}
\definecolor{tomato2}{rgb}{0.93,0.36,0.26}
\definecolor{tomato3}{rgb}{0.80,0.31,0.22}
\definecolor{tomato4}{rgb}{0.55,0.21,0.15}
\definecolor{tomato}{rgb}{1.00,0.39,0.28}
\definecolor{turquoise1}{rgb}{0.00,0.96,1.00}
\definecolor{turquoise2}{rgb}{0.00,0.90,0.93}
\definecolor{turquoise3}{rgb}{0.00,0.77,0.80}
\definecolor{turquoise4}{rgb}{0.00,0.53,0.55}
\definecolor{turquoise}{rgb}{0.25,0.88,0.82}
\definecolor{violetred}{rgb}{0.82,0.13,0.56}
\definecolor{violet}{rgb}{0.93,0.51,0.93}
\definecolor{wheat1}{rgb}{1.00,0.91,0.73}
\definecolor{wheat2}{rgb}{0.93,0.85,0.68}
\definecolor{wheat3}{rgb}{0.80,0.73,0.59}
\definecolor{wheat4}{rgb}{0.55,0.49,0.40}
\definecolor{wheat}{rgb}{0.96,0.87,0.70}
\definecolor{whitesmoke}{rgb}{0.96,0.96,0.96}
\definecolor{white}{rgb}{1.00,1.00,1.00}
\definecolor{yellow1}{rgb}{1.00,1.00,0.00}
\definecolor{yellow2}{rgb}{0.93,0.93,0.00}
\definecolor{yellow3}{rgb}{0.80,0.80,0.00}
\definecolor{yellow4}{rgb}{0.55,0.55,0.00}
\definecolor{yellowgreen}{rgb}{0.60,0.80,0.20}
\definecolor{yellow}{rgb}{1.00,1.00,0.00}

\newcommand{\arrayc}[1]{\ens{(c \times \cdots \times c)}}

\newcommand{\ens}[1]{\ensuremath{#1}}

\definecolor{nts}{rgb}{.8,.1,.8}
\definecolor{chk}{rgb}{.8,.2,.1}
\definecolor{reword}{rgb}{.9,.5,.7}

\newcommand{\nts}[1]{\textcolor{nts}{\ens{\rightarrow} #1 \ens{\leftarrow}}}
\renewcommand{\nts}[1]{ }
\newcommand{\chk}[1]{\textcolor{chk}{\ens{>~!} #1 \ens{!~<}}}
\renewcommand{\chk}[1]{ }

\newcommand{\imp}[1]{\textcolor{Red}{#1}}
\renewcommand{\imp}[1]{ }

\newcommand{\cons}[1]{\ens{C_{\mathsf{#1}}}}

\newcommand{\defcons}[2]{
	\ifbool{numsubs}{
		\edef#1{\noexpand\cons{\arabic{myc}}}
		\addtocounter{myc}{1}
	}
	{
		\newcommand{#1}{\cons{#2}}
	}
}

\newcommand*{\Rom}[1]{\expandafter\@slowromancap\romannumeral #1@}
\makeatother

\def\polylog{\operatorname{polylog}}

{\makeatletter
 \gdef\xxxmark{%
   \expandafter\ifx\csname @mpargs\endcsname\relax 
     \expandafter\ifx\csname @captype\endcsname\relax 
       \marginpar{xxx}
     \else
       xxx 
     \fi
   \else
     xxx 
   \fi}
 \gdef\xxx{\@ifnextchar[\xxx@lab\xxx@nolab}
 \long\gdef\xxx@lab[#1]#2{{\bf [\xxxmark #2 ---{\sc #1}]}}
 \long\gdef\xxx@nolab#1{{\bf [\xxxmark #1]}}
 \gdef\turnoffxxx{\long\gdef\xxx@lab[##1]##2{}\long\gdef\xxx@nolab##1{}}%
}

\newcommand{\ST}{\mathrm{ST}}
\newcommand{\SA}{\mathrm{SA}}

\newcommand{\ldotdot}{, \ldots, }
\newcommand{\abs}[1]{\lvert #1 \rvert}

\newcommand{\SCQ}{\mathrm{SCQ}}
\newcommand{\GSCQ}{\mathrm{GSCQ}}

\newcommand{\SKJ}{S[k \ldotdot j]}

\newcommand{\LCP}{\mathrm{LCP}}

\newcommand{\LZ}{\mathrm{LZ}}
\newcommand{\ILCP}{\mathrm{ILCP}}

\def\polylog{\operatorname{polylog}}

\sloppypar

\begin{document}

\author{
    Moshe Lewenstein\thanks{\texttt{moshe@cs.biu.ac.il}. This paper was written while on Sabbatical in U. of Waterloo. This research was supported by the U. of Waterloo and BSF grant 2010437, a Google Research Award and GIF grant 1147/2011.}
}

\institute{
    Bar-Ilan University
}

\title{Orthogonal Range Searching for Text Indexing}
\maketitle

\begin{abstract}
Text indexing, the problem in which one desires to preprocess a (usually large) text for future (shorter) queries, has been researched
ever since
the suffix tree was invented in the early 70's. With textual data continuing to increase and with changes in the way it is accessed,
new data structures and new algorithmic methods are continuously required. Therefore, text indexing  is of utmost importance
and is a very active research domain.

Orthogonal range searching, classically associated with the computational geometry community, is one of the tools that has
increasingly become important for various text indexing applications. Initially, in the mid 90's there were a couple of results
recognizing this connection. In the last few years we have seen an increase in use of this method and are reaching
a deeper understanding of the range searching uses for text indexing.

In this monograph we survey some of these results.
\end{abstract}

\tableofcontents
\newpage
\section{Introduction}
\label{sec:intro}

The {\it text indexing} problem assumes a (usually very large)
text that is to be preprocessed in a fashion that will allow
efficient future queries of the following type. A query is a
(significantly shorter) pattern. One wants to find all text
locations that match the pattern in time proportional to the {\sl
pattern length and number of occurrences}.

Two classical data structures that are most widespread amongst all
the data structures solving the text indexing problem are
the {\it suffix tree}~\cite{Weiner73} and the {\it suffix array}~\cite{MM93}
(see Section~\ref{sec:prepromdef} for definitions, time and space usage).

While text indexing for exact matches is a well studied problem, many other text indexing related
problems have become of interest as the field of text indexing expands.
For example, one may desire to find matches within subranges of the text~\cite{Makinen2006}, or to find which
documents of a collection contain a searched pattern~\cite{Muthukrishnan2002}, or one may want our text index compressed~\cite{nm-acs07}.

Also, the definition of a match may vary. We may be interested in a {\em parameterized} match~\cite{Baker96,Lewenstein08}, a {\em function} match~\cite{AALP06},
a {\em jumbled} match~\cite{AALS03,BEL04,CFL09,MR10} etc. These examples are only a very few of the many different interesting ways that the field
of text indexing has expanded.

New problems require more sophisticated ideas, new methods and new data structures. This indeed has happened in the realm
of text indexing. New data structures have been created and known data structures from other domains have been incorporated for the use
of text indexing data structures all mushrooming into an expanded, cohesive collection of
text indexing methods. One of these incorporated methods is that of orthogonal range searching
problems.

Orthogonal range searching refers to the preprocessing of a collection of points in $d$-dimensional
space to allow queries on ranges defined by rectangles whose sides are aligned with the
coordinate axes (orthogonal).

In the problems we consider here we assume that all input point sets are in rank space,
i.e., they have coordinates on the integer grid $[n]^d = \{0,\ldots , n-1 \}^d$. The rank-space
assumption can easily be made less restrictive, but we do not dwell on this here as the rank-space
assumption works well for most of the results here.

The set of problems one typically considers in range searching are queries on the range such as emptiness, reporting (all points in the
range), report any (one) point, range minimum/maximum, closest point. In general, some function on the set of points in the range.

We will consider different range searching variants in the upcoming sections and will discuss the time and space
complexity of each at the appropriate place. For those interested in further reading of orthogonal range
searching problems we suggest starting with~\cite{Agarwal97rangesearching,clp-socg11}.

Another set of orthogonal range searching problems is on arrays (not point sets). We will lightly discuss this
type of orthogonal range searching, specifically for Range Minimum Queries (RMQ).

In this monograph we take a look at some of the solutions to text indexing problems that have utilized range searching techniques. The reductions chosen are, purposely, quite straightforward with
the intention of introducing the simplicity of the use of this method. Also, it took some time for the pattern matching
 community to adopt this technique into their repertoire. Now more sophisticated reductions are emerging and
 members of the community have also been contributing to better range searching solutions, reductions for hardness and more.

\section{Problem Definitions and Preliminaries}\label{sec:prepromdef}

Given a string $S$, $\abs{S}$ is the length of $S$. Throughout this
paper we denote $n=\abs{S}$. An integer $i$ is a {\em location} or a
{\em position} in $S$ if $i = 1, \ldots, \abs{S}$. The substring
$S[i \ldotdot j]$ of $S$, for any two positions $i \leq j$, is
the substring of $S$ that begins at index $i$ and ends at index $j$.
The {\em suffix} $S_i$ of $S$ is the substring $S[i \ldotdot n]$.

\bigskip

\noindent
{\bf Suffix Tree}
The \emph{suffix
tree}~\cite{Weiner73,Ukkonen95,Farach-Colton2000,McCreight76} of a string
$S$, denoted $\ST(S)$, is a compact trie of all the suffixes of
$S\$$ (i.e., $S$ concatenated with a delimiter symbol $\$ \not\in
\Sigma$, where $\Sigma$ is the alphabet set, and for all $c\in\Sigma, \$ < c$). Each of its edges is labeled with a substring of $S$
(actually, a representation of it, e.g., the start location and its
length). The ``compact'' property is achieved by contracting nodes
having a single child. The children of every node are sorted in the
lexicographical order of the substrings on the edges leading to
them. Consequently, each leaf of the suffix tree represents a suffix
of $S$, and the leaves are sorted from left to right in the
lexicographical order of the suffixes that they represent.
$\ST(S)$ requires $O(n)$ space. The suffix tree can be prepared in $O(n+Sort(\Sigma))$,
where $n$ is the text size, $\Sigma$ is the alphabet, and $Sort(Q)$ is the time required to sort the
set $Q$~\cite{Farach-Colton2000}. For the suffix tree one can search an $m$-length pattern in $O(m+occ)$, where
$occ$ is the number of occurrences of the pattern. If the alphabet $\Sigma$ is large this potentially
increases to $O(m\log|\Sigma|+occ)$, as one need to find the correct edge exiting at every node. If
randomization is allowed then one can introduce hash functions at the nodes to obtain $O(m+occ)$,
even if the alphabet is large, without affecting the original $O(n+Sort(\Sigma))$
construction time.

\bigskip

\noindent
{\bf Suffix Array}
The \emph{suffix array}~\cite{MM93,Karkkainen2006} of a string $S$, denoted $\SA(S)$, is a permutation
of the indices $1,\ldots, n$ indicating the lexicographic ordering of the suffixes of $S$. For example,
consider $S = mississippi$. The suffix array of $S$ is
$[11, 8, 5, 2, 1, 10, 9, 7, 4, 6, 3]$, that is $S_{11} = "i" < S_8 = "ippi" < S_5 = "issippi" < \ldots < S_3 = "ssissippi"$, where $<$ denotes less-than lexicographically.
The construction time of a suffix array is $O(n+Sort(\Sigma))$~\cite{Karkkainen2006}. The time to answer an query
$P$ of length $m$ on the suffix array is $O(m + \log n + occ)$~\cite{MM93}\footnote{This requires LCP information. Details appear in Section~\ref{lcp-lemma}.}. The $O(m+\log n)$ is required to find the
range of suffixes (see Section~\ref{lcp-lemma} for details) which have $P$ as a prefix and then since $P$ appears as a prefix of suffix $S_i$ it must appear
at location $i$ of the string $S$. Hence, with a scan of the range we can report all occurrences in additional $O(occ)$ time.

\bigskip

\noindent
{\bf Relations between the Suffix Tree and Suffix Array}
Let $S = s_1s_2\ldots s_n$ be a string. Let $\SA = \SA(S)$ be its suffix array and $\ST = \ST(S)$ its suffix tree.
Consider $\ST$'s leaves. As these represent suffixes and they are in lexicographic ordering, $\ST$ is actually a tree over
$\SA$. In fact, one can even view $\ST$ as a search tree over $\SA$.

Say we have a pattern $P$ whose path from the root of $\ST$ ends on the edge entering node $v$ in $\ST$ (the locus).
Let $l(v)$ denote the leftmost leaf in the subtree of $v$ and $r(v)$ denote
the rightmost leaf in the subtree of $v$. Assume that $i$ is the location of $\SA$ that corresponds to $l(v)$,
i.e. the suffix $S_{\SA[i]}$ is associated with $l(v)$. Likewise assume $j$ corresponds to $r(v)$. Then
the range $[i,j]$ contains all the suffixes that begin with $P$ and it is maximal in the sense that
no other suffixes begin with $P$. We call this range the {\em $\SA$-range of $P$}.

Consider the previous example  $S = mississippi$ with suffix array
$[11, 8, 5, 2, 1, 10, 9, 7, 4, 6, 3]$. For a query pattern $P = si$ we have that the $\SA$-range
for $P$ is $[8,9]$, i.e. $P$ is a common prefix of
$\{s_{\SA[8]}\ldots s_n,s_{\SA[9]}\ldots s_n\} = \{sissippi, sippi\}$.

Beforehand, we pointed out that finding the $\SA$-range for a given $P$ takes $O(m + \log n)$ in the
suffix array. However, given the relationship between a node in the suffix tree and the SA-range in the suffix array, if we so desire,
we can use the suffix tree as a search tree for the suffix array and find the SA-range in $O(m)$ time. For simplification
of results, throughout this paper we assume that indeed we find $\SA$-ranges for strings of length $m$ in $O(m)$ time.

Moreover, one can find for $P=p_1,\ldots,p_m$ all nodes in a suffix tree representing
$p_i,\ldots, p_m$ for $1\leq i \leq m$ in $O(m)$ time using suffix links. Hence, one can find {\em all} $\SA$-ranges for
 $p_i,\ldots, p_m$ for $1\leq i \leq m$ in $O(m)$ time.

\section{1D Range Minimum Queries}
\label{sec:1d-rmq}

While the rest of this paper contains results for orthogonal range searching in rank space,
one cannot disregard a couple of important range searching results
that are widely used in text indexing structures. The range searching we refer to is the
{\em  Range Minimum Query} (RMQ) problem on an array (not a point set). RMQ is defined as follows.

Let $S$ be a set of linearly ordered elements whose elements can be compared (for $\leq$) in constant time.

\bigskip

\begin{tabular}{|l  l|}
  \hline
   \multicolumn{2}{|l|}{\bf $d$-Dimensional Range Minimum Query (d-RMQ)} \\
  \hline
{\bf \ \ Input:}& A d-dimensional array $A$ over $S$ of size $N = n_1 \cdot n_2 \cdot \ldots \cdot n_d$\ \ \ \ \ \\
 & where $n_i$ is the size of dimension $i$. \\
{\bf \ \ Output:\ \ }& A data structure over $A$ supporting the following queries.\\
{\bf \ \ Query:}& Return the minimum element in a range\\
&  $q = [a_1..b_1]\times [a_2..b_2]\times \ldots \times [a_d..b_d]$ of $A$.\\
  \hline
  \end{tabular}

\bigskip

1-dimensional RMQ plays an important role in text indexing data structures. Hence, we give a bit of detail on results about RMQ data structure construction.

The 1-dimensional RMQ problem has been well studied. Initially, Gabow, Bentley and Tarjan~\cite{gbt-stoc84} introduced the problem. They reduced the problem to the {\em Lowest Common Ancestor (LCA)} problem~\cite{ht-sjc84} on Cartesian Trees~\cite{Vuillemin1980}. The {\em Cartesian Tree} is a binary tree defined on top of an array of $n$ elements from a linear order. The root is the minimum element, say at location $i$ of the array. The left subtree is recursively defined as the Cartesian tree of the sub-array of locations $1$ to $i-1$ and the right subtree is defined likewise on the sub-array from $i+1$ to $n$. It is quite easy to see the connection between the RMQ problem and the Cartesian tree, which is what was utilized in~\cite{gbt-stoc84}, where the LCA problem was solved optimally in $O(n)$ time and $O(n)$ space while supporting $O(1)$ time queries. This, in turn, yielded the result of $O(n)$ preprocessing time and space for the 1D RMQ problem with answers in $O(1)$ time.

Sadakane~\cite{Sadakane07} proposed a position-only solution, i.e. one that return the position of the minimum rather than the minimum itself, of $4n + o(n)$ {\em bits} space with $O(1)$ query time.  
Fischer and Heun~\cite{FH:11} improved the space to $2n+o(n)$ bits and preprocessed in $O(n)$ time for subsequent $O(1)$ time queries. They also showed that the space must be of size $2n-O(\log n)$. 
Davoodi, Raman and Rao~\cite{DRS12} showed how to achieve the same succinct representation in a different way with $o(n)$ working space, as opposed to the $n+o(n)$ working space in~\cite{FH:11}.
It turns out that there are two different models, the {\em encoding} model and the {\em indexing model}. The model difference was already noted in~\cite{DL03}. For more discussion on the modeling differences see~\cite{bdr-algorithmica12}. In the encoding model we preprocess the array $A$ to create a data structure {\em enc} and queries have to be answered using {\em enc} only, {\em without} access to $A$. In the indexing model, we create an index {\em idx} and are able to refer to $A$ when answering queries. The result of Fischer and Heun~\cite{FH:11} is the encoding model result. For the indexing model Brodal et al.~\cite{bdr-algorithmica12} and Fischer and Heun~\cite{FH:11}, in parallel, showed that an index of size $O(n/g)$ {\em bits} is possible with query time $O(g)$. Brodal et al.~\cite{bdr-algorithmica12} showed that this is an optimal tradeoff in the indexing model.

Range minimum queries on an array have been extended to 2D in~\cite{Amir2007,ay-soda10,bdr-algorithmica12,bdlrr-esa12,Demaine2009,GIKRR:11} and to higher dimension $d$ in ~\cite{bdr-algorithmica12,cr-ijga91,ay-soda10,dll13}.

\subsection{The LCP Lemma}~\label{lcp-lemma}

The Longest Common Prefix (LCP) of two strings plays a very important role in text indexing and other string matching problems. So, define as follows.

\begin{definition}
Let $x$ and $y$ be two strings over an alphabet $\Sigma$. The {\em longest common prefix} of $x$ and $y$, denoted $LCP(x,y)$, is the largest string $z$ that is a prefix of both $x$ and $y$.
The length of $LCP(x,y)$ is denoted $|LCP(x,y)|$.
\end{definition}

The first sophisticated use of the LCP function for string matching was
for string matching with errors in a paper by Landau and Vishkin~\cite{LV88}. An interesting and very central result to text indexing structures appears in the following lemma, which is not
difficult to verify.

\begin{lemma}{\em ~\cite{MM93}}
Let $S^{[1]}, S^{[2]}, \ldots, S^{[n]}$ be a sequence of $n$ lexicographically ordered strings. Then ${\rm |LCP}(S^{[i]},S^{[j]})| = \min_{i\leq h < j} {\rm |LCP}(S^{[h]}, S^{[h+1]})|$.
\end{lemma}

This allows a data structure over the suffix array of size $O(n)$ that returns the LCP value of any two substrings in $O(1)$ time.
This is done by building an RMQ data structure over the array containing the values of the LCP of lexicographically consecutive suffixes and using the lemma.

This result was implicitly\footnote{They did not actually use the RMQ data structure. Rather, since they know the path a binary search will follow, they know which interval one needs to (RMQ)query when consulting a
given suffix array position (there is only one path towards it in the virtual binary search tree). So they directly store that range LCP value.} used in~\cite{MM93} to reduce the $O(m\log n)$ time for a search of an $m$-length pattern $P$ in a suffix array indexing a text of length
$n$ to $O(m+\log n)$. The idea is as follows. Both find the $\SA$-range for $P$ based on a binary search of the pattern $P$ on the suffixes of the suffix array.
The $O(m\log n)$ time follows for a naive binary search because it takes $O(m)$ time to check if $P$ is a prefix of a suffix and the $O(\log n)$ follows from the binary search.

Reducing to $O(m + \log n)$ is done as follows. The binary search is still used. Initially $P$ is compared to the string in the center of the lexicographic ordering.
This may take $O(m)$ time. However, at every stage of the binary search we maintain ${\rm |LCP}(P,T_i)|$ for the suffix $T_i$ of $T$ with the maximal ${\rm |LCP}(P,T_i)|$,
over all the suffixes to which $P$ has already been compared. When comparing $P$ to the next suffix, say $T_j$, in the binary
search, first ${\rm |LCP}(T_i,T_j)|$ is evaluated (in constant time) if ${\rm |LCP}(T_i,T_j)| \not= {\rm |LCP}(P,T_i)|$ we immediately know the value of
${\rm |LCP}(P,T_j)|$ - give it a moment of thought - and we can compare the character at location ${\rm |LCP}(P,T_j)|$+1 of $P$ and $T_j$ and continue the
binary search from there. Otherwise, ${\rm |LCP}(T_i,T_j)| = {\rm |LCP}(P,T_i)|$ in which case we continue the comparison of $P$ and $T_j$ (but only) from
the ${\rm |LCP}(P,T_j)|$+1-th character. Hence, one can claim, in an amortized sense, that the pattern is scanned only once. So, the search time is $O(m+\log n)$.

The dynamic version of this method is much more involved but has interesting applications, see~\cite{AFGKLL13}.

\subsection{Document Retrieval}~\label{Sec:Document-Retrieval}

The {\em Document Retrieval} problem is very close to the text indexing problem. Here we are given a collection of documents $D_1,\ldots, D_k$ and desire to preprocess them in order to answer document queries $Q$. A {\em document query} asks for the set of documents where $Q$ appears.

The {\em generalized suffix array} (for generalized suffix {\em tree}, see~\cite{Gusfield1997}) is a suffix array for a collection of texts $T_1, \ldots, T_k$ and can be viewed as the suffix array for $T_1\$_1T_2\$_2\ldots\$_{k-1}T_k$. However, we may remove, before finalizing the suffix array, all suffixes that start with a delimiter as they contain no interesting information. In order to solve the document retrieval problem one can build a generalized suffix array for $D_1,\ldots, D_k$. The problem is that when one seeks a query $Q$ one will find all the occurrences of $Q$ in all documents, whereas we desire to know only {\em in which} documents $Q$ appears and are not interested in all match locations.

A really neat trick to solve this problem was proposed by Muthukrishnan~\cite{Muthukrishnan2002}. Imagine the generalized suffix array for $D_1,\ldots, D_k$ of size $n=\sum_{1\leq i\leq k}|D_i|$ and a document retrieval query $Q$ of length $m$. In $O(m+\log n)$, or even in $O(m)$ time (as discussed in the end of Section~\ref{sec:prepromdef}) it is possible to find the $\SA$-range for $Q$. Now we'd like to report all documents who have a suffix in this range. So, create a {\em document array} for the suffix array. The {\em document array} for $D_1,\ldots, D_k$ will be of length $n$ and will contain at location $i$ the document id $d$ if $\SA[i]$ is a suffix beginning in document $d$. So, the former problem now becomes the problem of finding the unique id's in the $\SA$-range of the document array.

Muthukrishnan~\cite{Muthukrishnan2002} proposed a transformation to the RMQ problem in the following sense. Take the document array $DA$ and generate, yet another, array which we will call the {\em predecessor document array}. Let $\psi(i) = j$ if $j<i$, $DA(i)=DA(j)$ and for all $j<k<i, DA(k)\not=DA(i)$. $\psi(i) = -1$ if there is no such $j$. The predecessor document array has $\psi(i)$ at location $i$. The following observation now follows.

\begin{lemma}
Let $D_1,\ldots, D_k$ be a collection of documents and let $\SA$ be their generalized suffix array. Let $Q$ be a query and let $[i,j]$ be the $\SA$-range of $Q$. There is a one-one mapping between the documents in range $[i,j]$ in the document array and the values $< i$ in range $[i,j]$ in the predecessor document array.
\end{lemma}

\proof
Let $i_1 < i_2 < \ldots < i_r$ be all locations in $[i,j]$ where document id $d$ appears in the document array. Then the $i_1$-th location of the predecessor document array will be $< i$. However, locations $i_2 < i_3 < \ldots < i_r$ will contain $i_1, i_2, \ldots, i_{r-1}$, all greater than or equal to $i$, in the predecessor document array.
\qed

Hence, it is natural to consider an extended $RMQ$ problem defined now.

\bigskip

\begin{tabular}{|l  l|}
  \hline
   \multicolumn{2}{|l|}{\bf Bounded RMQ} \\
  \hline
{\bf \ \ Input:}& An array $A$.\ \ \ \ \ \\
{\bf \ \ Output:\ \ }& A data structure over $A$ supporting the following\\ & {\em bounded RMQ} queries.\\
{\bf \ \ Query:}& Given a range $[i,j]$ and a number $b$ find all values in the\\ & range $[i,j]$ of value $< b$.\\
  \hline
\end{tabular}

\bigskip

The bounded RMQ problem can be solved by recursively applying the known RMQ solution. Find an RMQ on $A[i,j]$, say it is at location $r$. If it is
less than $b$ then reiterate on $A[i,r-1]$ and $A[r+1,j]$. The preprocessing time and space are the same as those of
the RMQ problem. The query time is $O(ans)$, where $ans$ is the number of elements smaller than $b$.

This yields an $O(m + docc)$ solution for the document retrieval problem, where $docc$ is the number of documents in which the query $Q$ appears.

\section{Indexing with One Error}
\label{sec:oneError}

The problem of {\it approximate text indexing}, i.e. the text indexing problem where up to a given number of errors is
allowed in a match is a much more difficult problem than text indexing. The problem is formally defined as follows.

\bigskip
\noindent
\begin{tabular}{l l}
{\bf Input:} & Text $T$ of length $n$ over alphabet $\Sigma$ and an integer $k$.\\
{\bf Output:} & A data structure for $T$ supporting {\em $k$-error queries}.\\
{\bf Query:} & A $k$-error query is a pattern $Q = q_1q_2\ldots q_m$ of length $m$ over alphabet\\ & $\Sigma$ for which we desire to find all locations $i$ in $T$ where $Q$ matches\\ &  with $\leq k$ errors.
\end{tabular}
\bigskip

We note that there are several definitions of errors. The {\it
edit distance} allows for mismatches, insertions and
deletions~\cite{Levenshtein66}, the {\it Hamming distance} allows for
mismatches only. For text indexing with $k$ errors (for Hamming distance, Edit distance and more) Cole et al.~\cite{Cole2004} introduced a novel data
structure which, for the Hamming distance version, uses space $(n\log^kn)$ (it is preprocessed within an  $O(\log\log n)$ factor of the space complexity) and answers queries in $O(\log^kn + m + occ)$. See also~\cite{CLSTW11,Tsur10} for different space/time tradeoffs for the Hamming distance version. 


Throughout the rest of this section we focus and discuss the special case of one error. Moreover, we will do so for the
mismatch error, but a similar treatment will handle insertions
and deletions. The reduction to range queries presented in this section was obtained in parallel by Amir et al.~\cite{AKLLLR00} and
by Ferragina, Muthukrishnan and de Berg~\cite{FMD99}. The goal of~\cite{FMD99} was to show geometric data structures that solve certain
methods in object oriented programming. They also used their data structure to solve the dictionary matching with one error. In~\cite{AKLLLR00} Amir et al. solved
dictionary matching with one error and also solved the text indexing with one error. For the sake of simplicity, we
will present the result of text indexing with one error from~\cite{AKLLLR00}, but the reduction is the same for dictionary matching (see definition in~\cite{AKLLLR00}).

The algorithm that we will shortly describe combines a bidirectional construction of suffix trees, which had been known before.
Specifically, it is  similar to the data structure
of~\cite{BG96}. However, in~\cite{BG96} a reduction to 2D range searching was not used.

\subsection{Bidirectional Use of Suffix Arrays
\label{sec:bindex}}

For simplicity's sake we make the following assumption. {\sl Assume
that there are no {\bf exact} matches of the pattern in the text}. We
will relax this assumption later and show how to handle it in Section~\ref{sec:exact}.

{\bf The main idea:} Assume there is a pattern occurrence at text
location $i$ with a single mismatch in location $i+j-1$. This means that
$q_1 \ldots q_{j-1}$ has an {\sl exact match} at location $i$ and
$q_{j+1}\ldots q_m$ has an exact match at location $i+j$.

The distance between location $i$ and location $i+j$ is dependent on
the mismatch location, and that is somewhat problematic. We therefore
choose to ``wrap" the pattern around the mismatch. In other words, if
we stand exactly at location $i+j-1$ of the text and look left, we see
$q_{j-1} \ldots q_1$. If we look right we see $q_{j+1} \ldots q_m$.
This leads to the following algorithm.

\bigskip
For the data structure supporting $1$-mismatch queries construct a suffix array $\SA_{T}$ of text
string $T$ and a suffix array $\SA_{T^R}$ of the string $T^R$, where $T^R$
is the reversed text $T^R = t_n \ldots t_1$.

\noindent
In order to reply to the $1$-mismatch queries do as follows: 

\bigskip
\noindent
{\bf Query Reply:}
\begin{enumerate}
\item[\mbox{}] {\sf For $l=1,...,m$ do}
\item[1.] {\sf Find the maximal $\SA_T$-range $I_l = [i_l,j_l]$ of $q_{l+1}\ldots  q_m$ in
$\SA_T$, if it is non-empty.}
\item[2.] {\sf Find the maximal $\SA_{T^R}$-range $RI_l = [ri_l, rj_l]$ of $q_{l-1} \ldots q_1$ in
$\SA_{T^R}$, if it is non-empty.}
\item[3.] {\sf If both $I_l$ and $RI_l$ are non-empty, then return the intersection of $\SA_T$ and $\SA_{T^R}$ on their respective ranges.}
\end{enumerate}

\bigskip

Steps 1 and 2 of the query reply can be done for the $l$'s in {\it overall}
linear time (see end of Section~\ref{sec:prepromdef}). Hence, we only need an efficient implementation
of Step 3.

\begin{figure}[h!]
\centering
\includegraphics[scale=0.45]{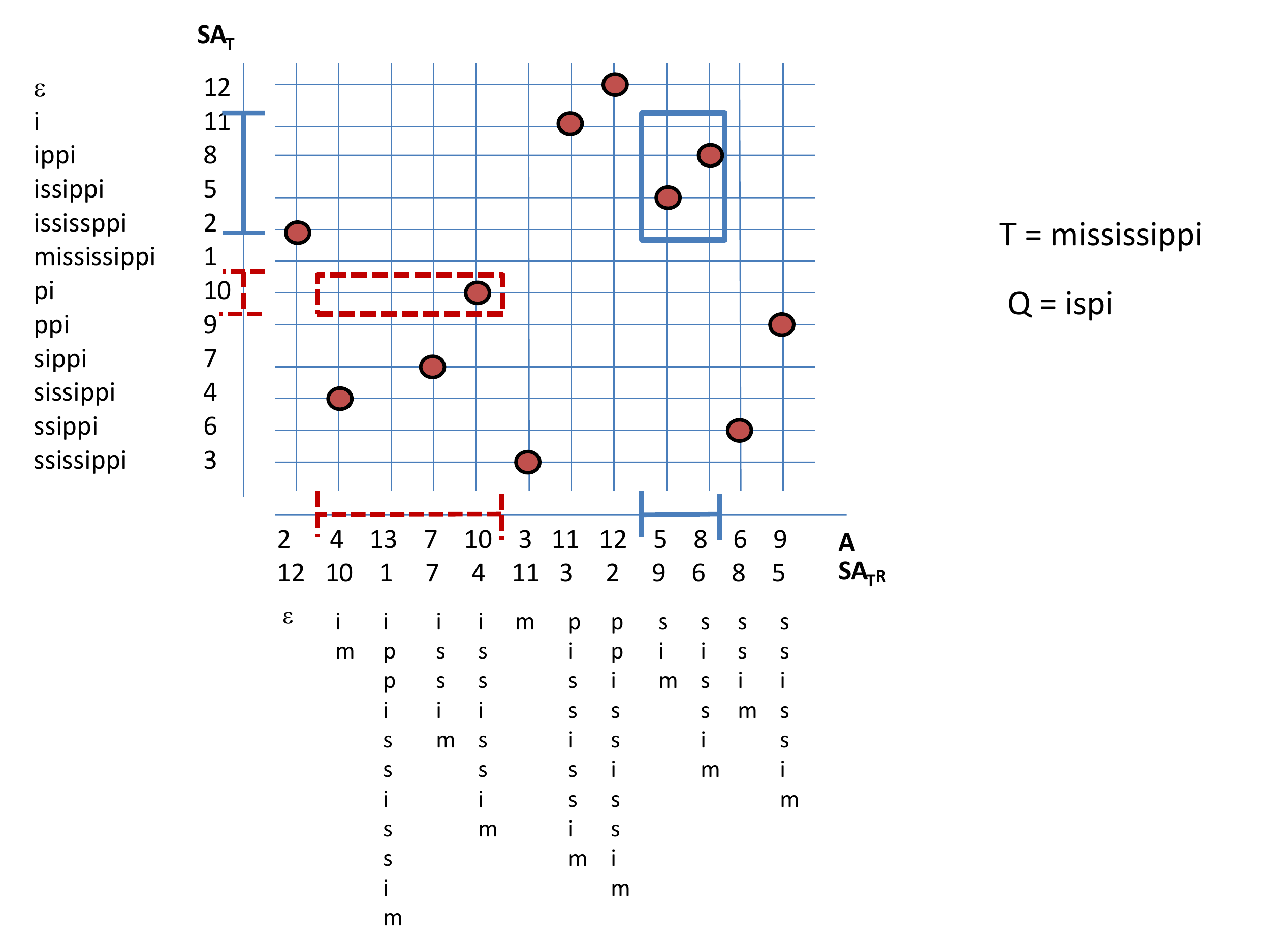}
\caption[]{A range query grid representing $T$ and $T^R$. The dashed rectangle represents a mismatch at location 2 of $Q$ and the solid rectangle represents a mismatch at location 3 of $Q$.}
\label{fig2}
\end{figure}

\subsection{Set Intersection via Range Reporting\label{sec:range}}

In Step 3, given $\SA$-ranges $I_l =$ [$i_l ... j_l$] and $RI_l =$ [$ri_l ... rj_l$] we want to report the points in the intersection of $SA_T$ and $SA_{T^R}$ w.r.t. to the corresponding coordinates of the two ranges. We show that this quite straightforwardly
reduces to the 2D range reporting problem.

\bigskip

\begin{tabular}{|l  l|}
  \hline
   \multicolumn{2}{|l|}{\bf Range Reporting in 2D (rank space)} \\
  \hline
{\bf \ \ Input:}& A point set $P = \{(x_1,y_1), \ldots, (x_n,y_n)\} \subseteq [1,n]\times[1,n]$.\ \ \ \ \ \\
{\bf \ \ Output:\ \ }& A data structure representing $P$ that supports the following\ \ \ \\ & {\em range reporting} queries.\\
{\bf \ \ Query:}& Given a range $R=[a,b]\times [c,d]$ report all points of $P$\ \ \ \\ & contained  in $R$.\\
  \hline
\end{tabular}

\bigskip

Since the arrays (the suffix array for $T$ and the suffix array for $T^R$) are permutations, every number between
1 and $n+1$ (we include suffix $\epsilon$) appears precisely once in each array. The coordinates of
every number $i$ are $(x_i,y_i)$, where $x_i=\SA_T^{-1}(i)$ and $y_i=SA_{T^R}^{-1}(n-i+3)$ (the choice of $n-i+3$ is to align the
appropriate reverse suffix with suffix $i$ --- explanation: $i$ becomes $n-i+1$ when reversing the text. Then one needs to move over one to the mismatch location and one more to the next location).
We define the point set to be $P=\{(x_2,y_2),\ldots ,(x_{n+1},y_{n+1})\}$
and construct a $2D$ range reporting structure for it (efficiency to be discussed in a moment). It is clear
that the range elements intersection corresponds precisely with a $2D$ query $I_l \times RI_l$.

The current best range reporting data structures in 2D are as follows:
\bigskip

\begin{enumerate}
\item
{\bf Alstrup, Brodal and Rauhe~\cite{ABR00}:} a data structure requiring $O(n\log^{\epsilon}n)$ space, for any constant $\epsilon$, that can answer queries in $O(\log\log n + k)$, where $k$ is the number of points reported.

\item
{\bf Chan, Larsen and P\v{a}tra\c{s}cu~\cite{clp-socg11}:} a data structure requiring $O(n\log\log n)$ space that can answer queries in $O(\log\log n(1 + k))$.

\item
{\bf Chan, Larsen and P\v{a}tra\c{s}cu~\cite{clp-socg11}:} a data structure requiring $O(n)$ space that can answer queries in $O(\log^{\epsilon}n(1 + k))$. Other succinct results of interest appear in a footnote\footnote{Note that prior succinct solutions show a novel adaptation of the method of Chazelle~\cite{chazelle-sjc88} to Wavelet Trees~\cite{GGV03}, see~\cite{Kar99}, \cite{Makinen2006} and~\cite{BCN-1201-3602}. It is especially worth reading the chapter of "Application as Grids" in~\cite{Navarro12} for more results along this line.}.
\end{enumerate}

Therefore, we have the following.

\begin{theorem}~\label{thm:oneError}
Let $T=t_1\ldots t_n$ and $Q=q_1\ldots q_m$. When no
exact match exists, indexing with one
error can be solved with $O(s(n))$ space such that queries can be answered in
$O(qt(n, m, occ))$ time, where:
\vspace{-0.3cm}
\begin{itemize}
\item[]
{\bf s(n)} =  the space for a range reporting data structure and
\item[]
{\bf occ} = the number of occurrences of $Q$ in $T$ with one error, and
\item[]
{\bf qt(n,m,occ)} =  the query time for the same range reporting data structure.
\end{itemize}
\end{theorem}

\noindent
{\bf Proof:}
 Other than the range reporting data structure the space required is $O(n)$. Likewise, Steps 1 and 2 of the query response require total time $O(m)$ for $j=1,...,m$. Hence, the space and time are 
 dominated by the range reporting data structure at hand, i.e. space $O(s(n))$ and query time $qt(n,m,occ)$.
$\qed$

\subsection{Indexing with One Error when Exact Matches Exist}
\label{sec:exact}

We assumed that the text contained no exact pattern occurrence in
the text. In fact, the algorithm would also work for the case where
there are exact pattern matches in the text, but its time complexity
would suffer. Recall that the main idea of the algorithm was to pivot a pattern position and check, for every text
location, whether the pattern to the left and to the right of the pivot were exact matches. However, if the pattern occurs as an exact match in the text then at that occurrence a match is announced for all $m$ pivots.
So, this means that every exact occurrence is reported $m$
times. The worst case could end up being as bad as $O(m*occ)$ (for
example if the text is $a^n$ and the pattern is $a^m$ then it would be $O(nm)$).

To handle the case of exact occurrences one can use the following
idea. Add a third dimension to the range reporting structure representing
the character in the text at the mismatch
location. The desired intersection is of all
suffix labels such that this character is {\sl
different from} the symbol at that respective pattern location. This leads to a specific variant of range searching.

\bigskip

\begin{tabular}{|l  l|}
  \hline
   \multicolumn{2}{|l|}{\bf 3D 5-Sided Range Reporting (rank space)} \\
  \hline
{\bf \ \ Input:}& A point set $P = \{(x_1,y_1, z_1), \ldots, (x_n,y_n, z_n)\}$\\ &  $\subseteq [1,n]\times[1,n]\times[1,n]$.\ \ \ \ \ \\
{\bf \ \ Output:\ \ }& A data structure representing $P$ that supports the\ \ \ \\ &  following {\em range reporting} queries.\\
{\bf \ \ Query:}& Given a range $R=[a,b]\times [c,d]\times [e,\infty]$ report all points\ \ \ \\ & of $P$ contained  in $R$.\\
  \hline
\end{tabular}

\bigskip

3D 5-Sided Range Reporting can be solved with space $O(n \log^{O(\epsilon)} n)$ and query time $O(occ+\log\log n)$~\cite{clp-socg11}.

Back to our problem. We need to update the preprocessing phase.

{\bf Preprocessing:} Preprocess for 3-dimensional range
queries on the matrix $[1,...,n] \times [1,...,n] \times
\Sigma$. If $\Sigma$ is unbounded, then use only the $O(n)$ symbols in
$T$. The new geometric points are $(x_i,y_i,z_i)$, where $x_i$ and $y_i$  will be the same as before
and $z_i = t_{i-1}$ will be the text character that needs to mismatch. This will be
added in the preprocessing stage.

The only necessary modification is for Step 3 of the query reply which becomes:
\begin{enumerate}
\item[3.] {\sf If $I_l$ and $RI_l$ both exist, then return all the points in
$I_l \times R_l \times \Sigma$ for which the z-coordinate is not the respective pattern mismatch symbol}.
\end{enumerate}

The above step can be implemented by two 3D 5-sided range queries on the three dimensional range $I_l \times R_l \times [1,a-1]$
and $I_l \times R_l \times [a+1,|\Sigma|]$ where $a$ is the current pattern symbol being examined. We assume that the
alphabet symbols are numbered $1,...,|\Sigma|$.

\begin{figure}[h!]
\centering
\includegraphics[scale=0.45]{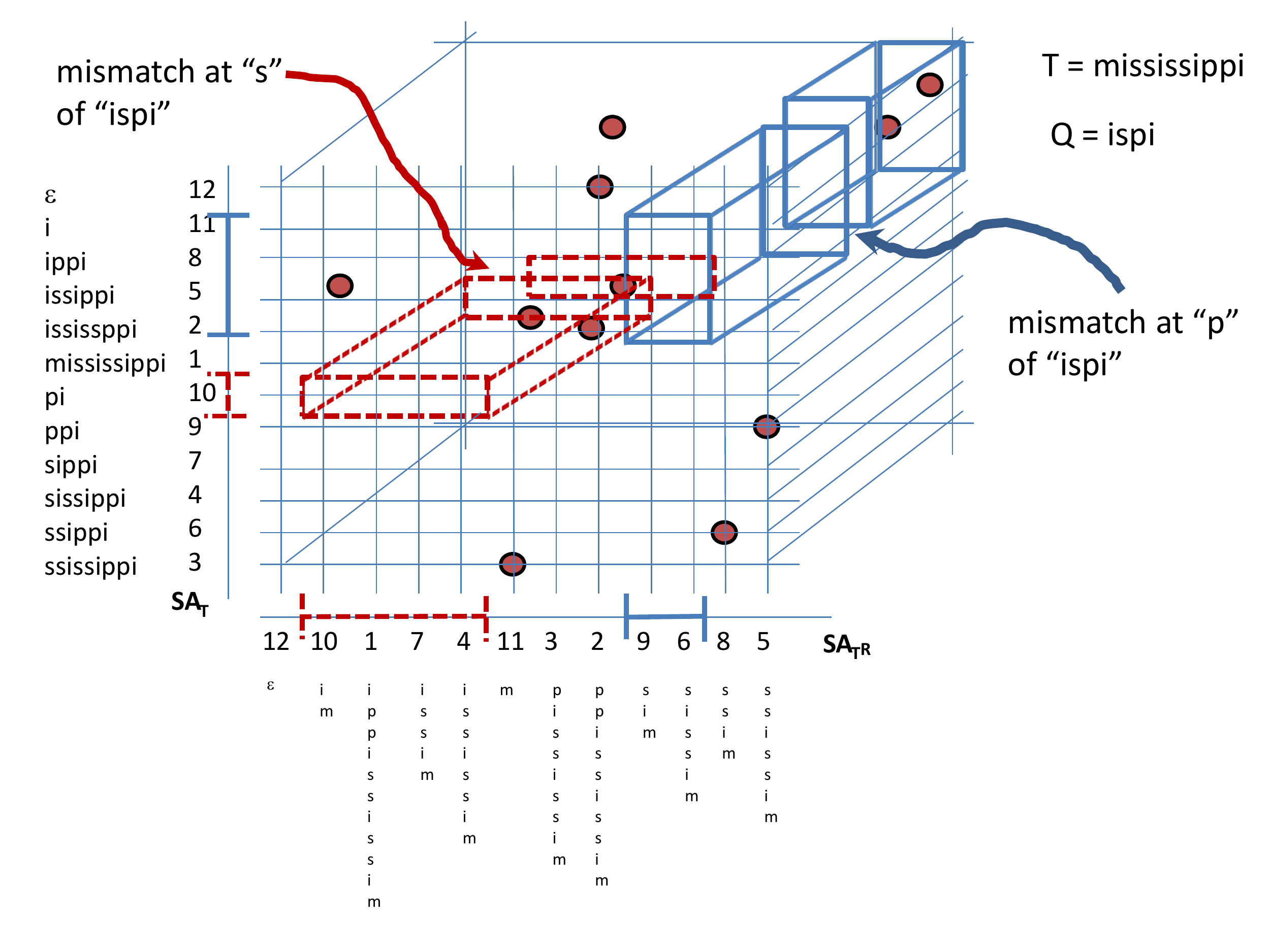}
\caption[]{The 3D 5-sided range queries on a 3D grid representing $T$ and $T^R$ and the appropriate "not to match" character. The dashed boxes represent a mismatch at location 2 of $Q$ and the solid boxes represent a mismatch at location 3 of $Q$.}
\label{fig-3d-half}
\end{figure}

See Figure~\ref{fig-3d-half} depicting the 3D queries.

\begin{theorem}\label{thm:atmost}
Let $T=t_1\ldots t_n$ and $Q=q_1\ldots q_m$.
Indexing with one error can be solved with $O(n \log^{O(\epsilon)} n)$
space and $O(occ + m \log\log n)$ query time,
where $occ$ is the number of occurrences of the pattern in the text
with at most one error.
\end{theorem}

{\bf Proof:} As in Theorem~\ref{thm:oneError}, the space and query time of the solution are dominated by the range searching structure. Hence, 
using the results of~\cite{clp-socg11}, the space is $O(n \log^{O(\epsilon)} n)$  and the query time is $O(occ +m\log\log n)$.
$\qed$

\subsection{Related Material}

For a succinct index for dictionary matching the best result appears in~\cite{HKSTV11}.

A {\em wildcard} character is one that matches all other symbols. When used we denote them with $\phi$.

Iliopoulos and Rahman~\cite{IR09} consider the problem of indexing a text $T$ to answer queries of the form $Q = Q_1\phi^dQ_2$, where $d, |Q_1|$ and $|Q_2|$ are known
during the preprocessing. The is known as gap-indexing. Bille and G{\o}rtz~\cite{BG11} consider the same problem. However, they required only $d$ to be known in advance. The solution in~\cite{BG11} uses a reduction
to range reporting. The reduction they apply is similar to the one presented in this chapter. In fact, it is easier because the gap's location within the query
pattern is known. So, one
does not need to check every position of the pattern as one does in the case of mismatch. Hence, the 2D range reporting data structure
is sufficient. One does need to adapt the search for a difference of
$d$ instead of $1$. This requires setting $y_i = n-i+d+2$.

\section{Compressed Full-Text Indexes} 
\label{sec:succinct-indexing}

Given that texts may be very large it makes sense to compress them - and there are many methods to do so. On the other hand,
these are not constructed to allow text indexing. Doing both at once has been the center of a lot of research activity
over the last decade. However, reaching this stage has happened in phases. Initially, pattern matching on compressed texts was considered starting by Amir et al.~\cite{ABF96}. By pattern matching on compressed texts we mean that a compressed text, with some predefined compressor, and a pattern are given and the goal is to find the occurrences of the pattern efficiently without decompressing the text. The second phase was text indexing while maintaining a copy of the original text and augmenting it with some sublinear data structure (usually based on a compressor) that would allow text indexing, e.g.~\cite{KU96}. We will
call this phase, the {\em intermediate phase}. Finally, compressed full-text indexing was achieved, that is text indexing without the original text
and with a data structure with size depending on the compressibility of the string. See Section~\ref{sec:1d-rmq} for the discussion on the encoding
model vs. the indexing model. Compressed full-text indexes of note are that of Ferragina and Manzini~\cite{FM05}, the {\em FM-Index}, that of Grossi and Vitter~\cite{GV05}, the {\em Compressed Suffix Array} and that of Grossi, Gupta and Vitter~\cite{GGV03}, the {\em Wavelet Tree}. For an extensive survey on compressed full-text indexes see~\cite{nm-acs07}.

In this section we will present two results, that of K{\"a}rkk{\"a}inen and Ukkonen~\cite{KU96} and that of Claude and Navarro~\cite{CN09}. The latter uses a central idea of the former. However, the former is from the intermediate phase. So, it maintains a text to reference. This certainly makes the indexing easier. The latter is a compressed full-text index. The former uses the more general LZ77 and the
latter uses SLPs (Straight Line Programs). Both LZ77 and SLP compressions are defined in this section.

\subsection{LZ77 Compressed Indexing}

The Lempel-Ziv compression schemes are among the best known and most widely used. In this section (and in Section~\ref{sec:substring-compression}) we will be interested
in the variant known as LZ77~\cite{ZL77}. For sake of completeness we describe the LZ77 scheme here.

\subsubsection{An Overview of the Lempel-Ziv Algorithm} Given an
input string $S$ of length $n$, the algorithm encodes the string in
a greedy manner from left to right. At each step of the algorithm,
suppose that we have already encoded $S[1 \ldotdot k-1]$ with $i-1$ phrases $\rho_1,\ldots, \rho_{i-1}$ (phrase - to be defined shortly). We search for
the location $t$, such that $1 \leq t \leq k-1$, for which the
longest common prefix of $S_k = S[k \ldotdot n]$ and the suffix $S_t$ is maximal.
Once we have found the desired location, suppose the aforementioned
longest common prefix is the substring $S[t \ldotdot r]$, a {\em phrase}, $\rho_i$,
will be added to the output which will include $F_i$ the encoding of the
distance to the substring (i.e., the value $k-t$), $L_i$ the length of
the substring (i.e., the value $r-t+1$), and the next character $C_i=S[k+(r-t+1)]$. The algorithm continues by encoding
$S_{k+(r-t+1)+1} = S[k+(r-t+1)+1 \ldotdot n]$. The sequence of phrases is called the string's (LZ77) {\em parse}
and is defined:

$$Z = \rho_1=(F_1, L_1,C_1),\rho_2=(F_2, L_2, C_2),\ldots, \rho_{c(n)}=(F_{c(n)}, L_{c(n)}, C_{c(n)})$$

We denote with $u(i) = \displaystyle\sum_{j=1}^{i-1}(L_j+1)+1$ the start location of $\rho_i$ in the string $S$. Finally, we denote
the output of the LZ77
algorithm on the input $S$ as $\LZ(S)$.

\subsubsection{K{\"a}rkk{\"a}inen and Ukkonen Method.}

Farach and Thorup~\cite{FT98}, in their paper on search in LZ77 compressed texts, noted a neat, useful observation.
Say $T$ of length $n$ is the text to be compressed and $Z$ is its LZ parse containing $c(n)$ phrases.

\begin{lemma}{\em \cite{FT98}}~\label{PrimaryOccurrences}
Let $Q$ be a query pattern and let $j$ be the smallest integer such that $Q = T[j\ldots j+|Q|-1]$. Then $u(i) \in [j, j+|Q|-1]$ for
some $1\leq i \leq c(n)$.
\end{lemma}

In other words, the first appearance of $Q$ in $T$ cannot be contained in a single phrase. K{\"a}rkk{\"a}inen and Ukkonen~\cite{KU96} utilized this
lemma to show how to augment the text with a sublinear text indexing data structure based on LZ77.

The idea is as follows. Say we search for a pattern $Q=q_1,\ldots, q_m$ in the text $T$ which has been compressed by LZ77. The occurrences of $Q$ in $T$
are defined differently for those that intersect with more than one phrase (which must exist if there are any matches according
to~Lemma~\ref{PrimaryOccurrences}) and those that are completely contained in a single phrase. The former are called {\em primary occurrences}
and the latter are called {\em secondary occurrences}.
The algorithm proposed in~\cite{KU96} first finds primary occurrences and then uses the
primary occurrences to find secondary occurrences.
Both steps use orthogonal range searching schemes.

In order to find the primary occurrences they used a bi-directional scheme based on Lemma~\ref{PrimaryOccurrences}.
Take every phrase $\rho_i = t_{u(i)},\ldots,t_{u(i+1)-1}$ and creates its reverse $\rho_i^R =  t_{u(i+1)-1},\ldots,t_{u(i)}$. Now,
consider a primary occurrence and the first phrase it starts in, say $\rho_i$. Then there must be a $j$ such that $q_1,\ldots, q_j$ is
a suffix of $\rho_i$ and $q_{j+1},\ldots, q_m$ is a prefix of the suffix of $T$, $t_{u(i+1)}, \ldots, t_n$. So, to find the
primary occurrences it is sufficient to find $k$ such that:

\begin{enumerate}
\item
$q_j,\ldots, q_1$ is a prefix of $\rho_i^R$.
\item
$q_{j+1},\ldots, q_m$ is a prefix of the suffix $t_{u(i+1)}, \ldots, t_n$ of $T$.
\end{enumerate}

Let $\pi$ be the lexicographic sort of $\rho_1^R, \rho_2^R,\ldots, \rho_{c(n)}^R$, i.e. $\rho_{\pi(1)}^R < \rho_{\pi(2)}^R < \ldots < \rho_{\pi(c(n))}^R$.
This leads to a range reporting scheme, similar to that of the previous section, of size $c(n)\times c(n)$, where the point set $P = \{(x(i),y(i)\ |\ x(i)=\SA_T^{-1}(u(i+1)), y(i)=\pi^{-1}(i)\}$.

Therefore, for every $1\leq j \leq m$ we need to (1) find the range of $\pi$ for which every $\rho_{\pi(i)}^R$ within has $q_j,\ldots, q_1$ as a prefix and (2) find the range
of suffixes that have $q_{j+1},\ldots, q_m$ as a prefix. Once we do so we revert to range reporting, as in the previous section. However, finding the ranges is not as simple as
in the previous section. Recall that we want to maintain the auxiliary data in sublinear space. So, saving a suffix array for the text is not a possibility. Also, the reversed phrases need to be indexed for finding the range of relevant phrases.

To this end a sparse suffix tree was used in~\cite{KU96a}. A {\em sparse suffix tree} is a
compressed trie over a subset of the suffixes and is also known as a {\em Patricia Trie} over this suffix set.
As it is a compressed trie it is of size $O($subset size$)$, in our case $O(c(n))$. Lately, in~\cite{BFGKSV13} it was shown how to construct a sparse suffix
array in optimal space and near-optimal time.

The sparse suffix tree can be constructed for the suffixes $T_{u(i+1)}$ and, since we have the original text on hand, we can maintain the compressed suffix tree
in $O(c(n))$ words. Navigation on this tree is the same as in a standard suffix tree. For the reversed phrases we can associate each with a prefix of the text. Reversing them
gives a collection of suffixes of the reversed text. Now construct a sparse suffix tree for these reversed suffixes and this will allow finding the range in $\pi$ (with the help of the
existing text) as in the previous section. Hence,

\begin{theorem}
Let $T$ be a text and $Z$ its LZ77 parse. One can construct a text indexing scheme which maintains $T$ along with a data structure of size $O(|Z|)$ such that for a query $Q$ we can find all its primary occurrences in $O(|Q|(|Q|+\log^{\epsilon}|Z|) + occ*\log^{\epsilon}|Z|)$, where $occ$ is the number of primary occurrences.
\end{theorem}

\proof
For the query $Q$, as described above, each pattern position is evaluated for primary occurrences. That is for each of the $|Q|$ pattern positions, we first traverse the auxiliary sparse suffix trees in $O(|Q|)$ time and once the ranges (for that pattern position is found) we perform a range query. The traversal will cost $O(|Q|^2)$ time.

Using the 2D succinct range reporting of Chan et al.~\cite{clp-socg11} (see previous section) we have linear space, i.e. $O(|Z|)$, and $O(\log^{\epsilon}|Z|(1+k))$ query time where $k$ is the number of points found, which is the same as the number of primary occurrences found.

Hence, over the $|Q|$ pattern positions the range querying will cost $\sum_{i=1}^{|Q|} \log^{\epsilon}|Z|(1+occ_i)$ time, where $occ_i$ is the number of primary occurrences when we split the pattern at position $i$. However, $\sum_{i=1}^{|Q|} \log^{\epsilon}|Z|(1+occ_i) = \sum_{i=1}^{|Q|} \log^{\epsilon}|Z| + \sum_{i=1}^{|Q|} \log^{\epsilon}|Z|*occ_i = |Q|\log^{\epsilon}|Z| + \log^{\epsilon}|Z|occ$. Recalling that the traversal cost $O(|Q|^2)$ time yields the desired.
\qed

\bigskip

We note that the secondary occurrences still need to be found. This is another interesting part of the paper and we refer the interested reader to~\cite{KU96}.

See the following papers for more along the following line using Lempel Ziv compressors~\cite{ANS12,FM05,KN13,RO08}.

\subsection{SLP Text Indexing}

Claude and Navarro~\cite{CN09} proposed a full-text indexing scheme based on {\em straight line programs} ({\em SLP}s). An {\em SLP} is a grammar
based compression for a text $T$. The grammar produces
exactly one word $T$ and the rules are in Chomsky Normal Form, i.e. each rule is $A \rightarrow BC$, where $A,B,C$ are variables of the grammar or $A\rightarrow a$,
where $A$ is a variable and $a$ is a terminal (a character of $T$).

The text indexing scheme that they propose follows the previous idea~\cite{KU96} of finding primary and secondary occurrences.
However, for SLPs things are slightly different. Consider the derivation
tree for the text $T$, that is deriving the full word $T$ by generating from the start symbol $S$ as the root (if $S\rightarrow AB$ then $A$ and $B$ will be children of the root
$S$ in the derivation tree - from here the derivation tree is applied recursively until the full $T$ is spelled out in the left-to-right order of the leaves). Every occurrence of
a pattern $Q$ in the text $T$ has a unique lowest variable $V$ which produces this occurrence, but its children do not. That is if the children of $V$ are $V_1$ and $V_2$, i.e. there
is a rule $V\rightarrow V_1V_2$, then $V_1$ produces $x, q_1, \ldots, q_j$ (where $x\in\Sigma^*$) and $V_2$ produces $q_{j+1}, \ldots, q_m, y$ (where $y\in\Sigma^*$) for some $j$. We say that $V$ splits pattern $Q$ at location $j$. An occurrence of $Q$ is called a {\em primary occurrence} if for some $V$ and $j$ $V$ splits this occurrence of $Q$. All other occurrences are {\em secondary occurrences}.

The format of the algorithm is to, once again, find the primary occurrences and then to deduce the occurrences of $Q$ in the text therefrom. With the goal of finding the primary occurrences in mind, once again, our grid will be of size $c(n) \times c(n)$, where $c(n)$ is the size of the variable set of the grammar. Each side of the grid will have one coordinate for each variable. The range searching point set is defined per rule, $V\rightarrow V_1V_2$. The location $(x,y)$ on the grid corresponding to $V_1$ (for $x$) on one side and $V_2$ (for $y$) on the other will have a point, $V$, on the grid. The ordering of the variables on either side of the grid follows from the desire to satisfy the following conditions.

\begin{enumerate}
\item
$q_j,\ldots, q_1$ is a prefix of $V_i^R$.
\item
$q_{j+1},\ldots, q_m$ is a prefix of $V_l$.
\item
There is a rule $V\rightarrow V_iV_l$.
\end{enumerate}

It is easy to see that the desired ordering, as in the LZ77 scheme, has the phrases in the $x$-coordinate in reverse lexicographic ordering and has the phrases in the $y$-coordinate in 
lexicographic ordering. The challenge here is to actually find the range of variables where $q_j,\ldots, q_1$ is a prefix of $V_i^R$. This is because it is a full-text index and
the text is not accessible any more. Nevertheless, this is doable in the SLP compression scheme using a suffix array type of search and comparing $q_j,\ldots, q_1$ with the variable at
hand. This comparison is not trivial. However, the full scheme is out of scope of this survey and we refer the reader to the full paper~\cite{CN09}. The result achieved is as follows:

\begin{theorem}
Let $T$ be a text of size $N$ represented by an SLP with $n$ variables and height $h$. There is a representation using $n(\log N+ 3 \log n+O(\log |\Sigma|+ \log h) +o(logn))$ bits such that $Q$ of length can be found in $O((m(m+h)+h occ)\log n)$ query time.
\end{theorem}

An extension of this idea to a general grammar can be found in~\cite{CN12a}, where the dependency on $h$ was removed from the search time. There is also other work for different compressors. See~\cite{GGKNP12} for one of the latest.

\section{Weighted Ancestors} 
\label{sec:SARange}

\subsection{2-Sided Sorted Range Reporting in 2D}

In this section we consider the {\em 2-sided sorted range reporting} problem\footnote{Results in Section 6.1 stem from wonderful research chats with Timothy Chan.} which is defined now.

\bigskip

\begin{tabular}{|l  l|}
  \hline
   \multicolumn{2}{|l|}{\bf 2-Sided Sorted Range Reporting in 2D} \\
  \hline
{\bf \ \ Input:}& A point set $P = \{(x_1,y_1), \ldots, (x_n,y_n)\} \subseteq [1,U]\times[1,U]$.\ \ \ \ \ \\
{\bf \ \ Output:\ \ }& A data structure representing $P$ that supports the following\ \ \ \\ & {\em 2-sided sorted range reporting} queries.\\
{\bf \ \ Query:}& Given a range $R=[-\infty, a]\times [-\infty, b]$ report all points of $P$\ \ \ \\ & contained  in $R$ sorted by their $y$-coordinate\\ &  (from highest to lowest).\\
  \hline
\end{tabular}

\bigskip
Note that we deviate from the assumption that the points are in rank-space. This is important for the the application of this section. 
We now show a method to solve the 2-sided sorted range reporting. The idea is as follows.

Consider the dynamic predecessor problem in which we need to support the following operations (a) insertions/deletions of integers ($\in [1,U]$) and (b) predecessor queries. This is a classical problem and is solved with a van-Emde Boas tree~\cite{veBoas77} or with $y$-fast tries~\cite{Willard83} in $O(n)$ space ($n$ is the current number of integers) and $O(\log\log U)$ time for the operations, where $[1,U]$ is the domain of the elements.

Dietz and Raman~\cite{DR91} asked whether this could be made partially persistent\footnote{Actually Dietz and Raman~\cite{DR91} asked about persistency in general, which may refer to full persistence or partial persistence. We stick to partial persistence as it is sufficient for our needs.} within the same query times. In other words can one create a data structure where insertions and deletions are supported on the current version but predecessor queries can be made on any of the versions (current or previous) of the data structure. Recently, Chan~\cite{Chan11} accomplished this by constructing a partially persistent predecessor data structure with space $O(n)$ and operations time $O(\log\log U)$. Chan's result~\cite{Chan11} is in fact more general, showing that the first predecessor can be found (in any previous version) in $O(\log\log U)$ time but the predecessor of the predecessor (etc.) can be found in $O(1)$ time. This yields a time of $O(\log\log U + k)$ to find the $k$ previous elements in sorted order in a chosen version of the data structure.

We utilize this for the 2-sided sorted range reporting by creating a data structure for $P$ as follows\footnote{We point out that for our purposes, finding one successor, the results of Dietz and Raman~\cite{DR91} are sufficient because (a) we seek only one successor and (2) the insertions are done first and then the queries are asked.}. Consider the sort of the $x$-coordinates of $P$, i.e. some permutation $\pi$ for which $x_{\pi(1)} < x_{\pi(2)} < \ldots < x_{\pi(n)}$. Now we insert the $y$-coordinates into the data structure according to $\pi$. That is we insert $y_{\pi(1)}$ and then $y_{\pi(2)}$ until $y_{\pi(n)}$. Now, a 2-sided sorted range reporting query $R=[-\infty, a]\times [-\infty, b]$ is answered as follows; first use a predecessor query to find $a$ within $x_{\pi(1)}, x_{\pi(2)}, \ldots, x_{\pi(n)}$ - that is find $i$ such that $x_{\pi(i)}\leq a < x_{\pi(i+1)}$. Then we go to the $i$-th copy of the partially persistent data structure which contains the points $(x_{\pi(1)},y_{\pi(1)}), (x_{\pi(2)},y_{\pi(2)}), \ldots, (x_{\pi(i)},y_{\pi(i)})$. Hence, the points are exactly the points that satisfy that their $x$-coordinate $\in [-\infty, a]$. Now to find the relevant points ($y\in [-\infty, b]$) in $R$ sorted by their $y$-coordinate we need to apply the predecessor query. This yields an $O(\log\log U + occ)$ when using the data structure from~\cite{Chan11}. Hence,

\begin{theorem}
The 2-sided sorted range reporting problem on an $n$-point set over a $U\times U$ grid can be solved with $O(n)$ space and $O(\log\log U + occ)$ time.
\end{theorem}

\subsection{Weighted Ancestors to 2-Sided Range Successor in 2D}

Consider the weighted ancestors problem on an edge-weighted tree introduced by Farach and Muthukrishnan~\cite{FM96} for the sake of obtaining a perfect-hash for substrings. An edge-weighted tree is a tree where each edge $e$ has a weight $w(e) \in [1,U]$. Each node $v$ is associated with a weight $w(v) = \sum_{e\in p_v}w(e)$, where $p_v$ is the path from root-to-$v$. The {\em weighted ancestors} problem is defined as follows.

\bigskip

\noindent
{\bf Input:} An edge-weighed tree $T$ with weight function $w$.

\noindent
{\bf Ouput:} A data structure supporting {\em weighted ancestor queries}.

\noindent
{\bf Query:} Given a node $u$ and a threshold $t$ find the ancestor $v$ of $u$ such that $w(v) \geq t$, but $w(p(v)) < t$, where $p(v)$ is the parent of $v$.

\bigskip
The weighted ancestor problem is a natural extension of the predecessor problem to trees. The application considered by~\cite{FM96} was on suffix trees. A suffix tree can be viewed as an edge-weighted tree with the edge weights denoting the length of the text with which the edge is marked. Now, say you are given indices $i$ and $j$ and want to find the locus of $T[i\ldots j]$ in the suffix tree. This can be done by going to the leaf representing $i$ and asking a weighted ancestor query with threshold $j-i+1$. The answer to the query is the locus of $T[i\ldots j]$.

In~\cite{FM96} a solution was given with $O(n\log n)$ preprocessing time, $O(n)$ space, and $O(\log\log U)$ query time. Their solution is based on a heavy path decomposition in order to linearize the input tree. Each path of the heavy path decomposition is assigned a predecessor structure. The preprocessing time of $O(n\log n)$ can be improved to $O(n)$ and this has been pointed out in~\cite{ALLS07,KL07}. It should be mentioned that the authors of~\cite{FM96} were considering a PRAM model and hence the $O(n\log n)$ time is really $O(\log n)$ parallel time and $O(n)$ work. Lately, it was shown that if the depth of the answer is $d$ in the tree then the query can be answered in $O(\log\log d)$ time~\cite{KNS12}.

We now present a solution for this problem using 2-sided sorted range reporting.

Consider the edge-weighted input tree $T$. We assume that every internal node in the tree has at least two children. Otherwise, create a dummy child with an arbitrary edge weight, say 1. Now consider the leaves $\l_1,\ldots, l_k$ ordered in inorder. For every two adjacent leaves denote their lowest common ancestor with $LCA(l_i, l_{i+1})$ and $nw(i) = w(LCA(l_i, l_{i+1}))$. With one scan of the tree all these values are computable. Now generate an array of the $nw$ values. Say we are given a weighted ancestor query, node $u$ and threshold $t$. We may assume that $u$ is a leaf. Otherwise, we simply choose a descendant leaf to represent $u$ (the answer will be the same). Consider the node $v$ which is the answer to the query and consider its parent $p(v)$. Since $p(v)$ has at least two children $v$ has at least one sibling. Say, $v$ has a sibling to its left (in inorder). Let $u=l_i$ then in the $nw$ array the first location $j < i$ that satisfies $nw(j) < t$ is the node for which the $LCA(l_j,l_i)=p(v)$. To obtain this $j$ we revert to 2-sided sorted range reporting. We set the points on the grid to be $P = (j,nw(j))$. The query is bounded by $i$ in the $x$-coordinates and $t$ in the $y$-coordinates. What we are looking for is the first answer, the element with the largest $y$-coordinate.

Note that once this is done it is still necessary to find $v$ (we only obtained $p(v)$). This can be done with a predecessor structure for each node. That is, for each node $p(v)$ save the index $h$ of the leftmost leaf $l_h$ for each of $p(v)$'s children. A predecessor query with $i$ will return the correct edge with child $v$, the weighted ancestor of $u$. Hence, 

\begin{theorem}
Let $T$ be an $n$ node edge-weighted tree with weights from $[1,U]$. Then using 2-sided sorted range reporting one can answer weighted ancestor queries in $O(\log\log U)$ time. The space required is $O(n)$. 
\end{theorem}

Note that for a suffix tree, the motivation in~\cite{FM96}, the weights are from $[1,n]$. So, the query time for a suffix tree is $O(\log\log n)$.

Recall (from Section~\ref{sec:1d-rmq}) the definition of an $\SA$-range and its relation to the suffix tree. Hence, the method just described precisely finds the boundaries of the $\SA$-range for a given suffix $i$ (in the suffix array) and its prefix of length (threshold) $t$.

\section{Compressed Substring Retrieval} 
\label{sec:substring-compression}

In this section we are concerned with the {\em substring compression} problem. 
The {\em substring compression} problem was introduced in~\cite{CM05}. Some of the definitions and layout here are from~\cite{CM05}. The solution, specifically the reduction to {\em range successor queries}, is from~\cite{KKLL09}.

In {\em substring compression} one is given a text to preprocess so that,
upon request, a compressed substring is returned. The goal is to do so quickly,
preferably in $O(c(s))$ time, where $c(s)$ is the size of the compressed substring.
{\em Generalized substring compression} is the same
with the following twist. The queries contain an additional context
substring (or a collection of context substrings) and the answers
are the substring in compressed format, where the context substring
is used to make the compression more efficient.

The compressor of interest is, once again, LZ77. We use the terminology from Section~\ref{sec:succinct-indexing}.
Some extra terminology is as follows. The string $S$ may be encoded within the context of the string $T$.
We denote this by $\LZ(S\mid T)$. The encoded result will be equivalent to the result when LZ77 is performed on the
concatenated string $T\$S$, where \$ is a symbol that does not
appear in either $S$ or $T$. However, only the portion of
$\LZ(T\$S)$ which represents the compression of $S$ is output by
the algorithm.

Formally, given a string $S$ of length $n$, we wish to preprocess $S$ in such
a way that allows us to efficiently answer the following queries:

\begin{description}
\item[Substring Compression Query ($\SCQ(i,j)$):]
given any two indices $i$ and $j$, such that $1 \leq i \leq j \leq
n$, we wish to output $\LZ(S[i \ldotdot j])$.
\item[Generalized Substring Compression Query ($\GSCQ(i,j,\alpha,\beta)$):]
given any four indices $i$, $j$, $\alpha$, and $\beta$, such that $1
\leq i \leq j \leq n$ and $1 \leq \alpha \leq \beta \leq n$, we wish
to output $\LZ(S[i \ldotdot j]\mid S[\alpha \ldotdot \beta])$.
\end{description}

The goal is to do answer queries quickly. Query times for both of the above query types will strongly
depend on the number of phrases actually encoded. We denote these
as $C(i,j)$ and $C_{\alpha,\beta}(i,j)$ for SCQ and GSCQ,
respectively.

\subsection{SCQ to Range Successor in 2D}

Recall the definition of LZ77 from Section~\ref{sec:succinct-indexing}. Imagine that we have already computed the phrases for $S[i\ldots k-1]$ and desire to compute the next phrase which is a prefix of $S[k\ldots j]$. In other words, we want to find the location $i \leq t \leq k-1$ for which the longest common
prefix of $\SKJ$ and the suffix $S_t$ is maximal. Consider the suffix $S_k$, which is an extension of $S[k\ldots j]$. Clearly, it is sufficient to find the
suffix $S_t$ for which $\abs{\LCP(S_k,S_t)}$ is maximized (without necessarily computing the value $\abs{\LCP(S_k,S_t)}$ at this stage).
Therefore we have two steps:
(1) finding the location $t$, and (2) computing
$\abs{\LCP(S_k,S_t)}$. Step (2) is easy since we assume that we have a full LCP data structure as described in Section~\ref{sec:1d-rmq}. So, our goal is to solve Step (1).
To do so we generalize our problem to the following.

\begin{description}
\item[Interval Longest Common Prefix ($\ILCP(k,l,r)$):] given $k,l,r$, we look for location $l \leq t \leq r$ of $S$ for which the suffix $S_t$ has the longest common prefix with $S_k$.
\end{description}

\noindent
Clearly, for us it is sufficient to compute $\ILCP(k,i,k-1)$. 

To compute the Interval Longest Common Prefix ($\ILCP(k,l,r)$) we use a reduction to the problem of {\em 3-sided range successor query}. That is given a 2D rank-space input on an $n\times n$ grid, a 3-sided query $R=[a, b]\times [-\infty, c]$ seeks the point in $R$ with the largest $y$-coordinate. The 3-sided range successor query problem was considered under a different guise in~\cite{CIKRTW12} where it was called the {\em range next value} problem. There it was considered as an array problem for which one desires to preprocess the array to allow queries that seek the largest value on a range less than a value $v$. This can be translated to a grid and vice versa.

\subsection{4-Sided and 3-Sided Sorted Range Reporting}\label{sec:Finding}

 The 3-sided range successor query generalizes quite nicely to the {\em 3-sided sorted range reporting in 2D} which we define now.

\bigskip

\begin{tabular}{|l  l|}
  \hline
   \multicolumn{2}{|l|}{\bf 3-Sided Sorted Range Reporting in 2D (rank space)} \\
  \hline
{\bf \ \ Input:}& A point set $P = \{(x_1,y_1), \ldots, (x_n,y_n)\} \subseteq [1,n]\times[1,n]$.\ \ \ \ \ \\
{\bf \ \ Output:\ \ }& A data structure representing $P$ that supports the following\ \ \ \\ & {\em 3-sided sorted range reporting} queries.\\
{\bf \ \ Query:}& Given a range $R=[a, b]\times [-\infty, c]$ report all points of $P$\ \ \ \\ & contained  in $R$ sorted by their $y$-coordinate\\ &  (from highest to lowest).\\
  \hline
\end{tabular}

\bigskip

A solution for the 3-sided range successor query problem was proposed in Lenhof and Smid~\cite{LS94}\footnote{Note, they called it the {\em Range Searching for Minimum} problem.}, and modified in~\cite{KKL07} (improved query times) to work in rank space, i.e. on an $[n] \times [n] $ grid for $n$ values
with queries supported in $O(\log \log n)$ worst-case time, using $O(n \log n)$ space. However, there are now better results which solve, not only the 3-sided range successor problem, but the more general 3-sided sorted range reporting.  The 3-sided sorted range reporting generalizes the 3-sided range successor query because the solutions presented can report the first location and stop. 


For the same reason the 3-sided solutions work just as well for the 4-sided sorted range reporting, as they can report all points until the $y$-coordinates surpasses the range boundary and then stop.

The current best range 3-sided sorted range reporting data structures in 2D are as follows:

\begin{enumerate}
\item
{\bf Navarro and Nekrich~\cite{NN12}:}\\ {\bf (a):} a data structure with $O(n)$ space where queries can be answered in $O(\log n(1 + k))$ where $k$ is the number of points reported and\\ {\bf (b):} a data structure with $O(n\log\log n)$ space where queries can be answered in $O(\log\log n(1 + k))$ and\\
{\bf (c):} a data structure with $O(n\log^{\epsilon}n)$ space for any constant $\epsilon>0$, where queries can be answered in $O(\log\log n + k)$, 

\item
{\bf Crochemore et al.~\cite{CIKRTW12}:} a data structure that requires $O(n^{1+\epsilon})$ space for any constant $\epsilon > 0$ and can answer queries in $O(k)$ time.
\end{enumerate}



\begin{figure}[h!]
\centering
\includegraphics[scale=0.5]{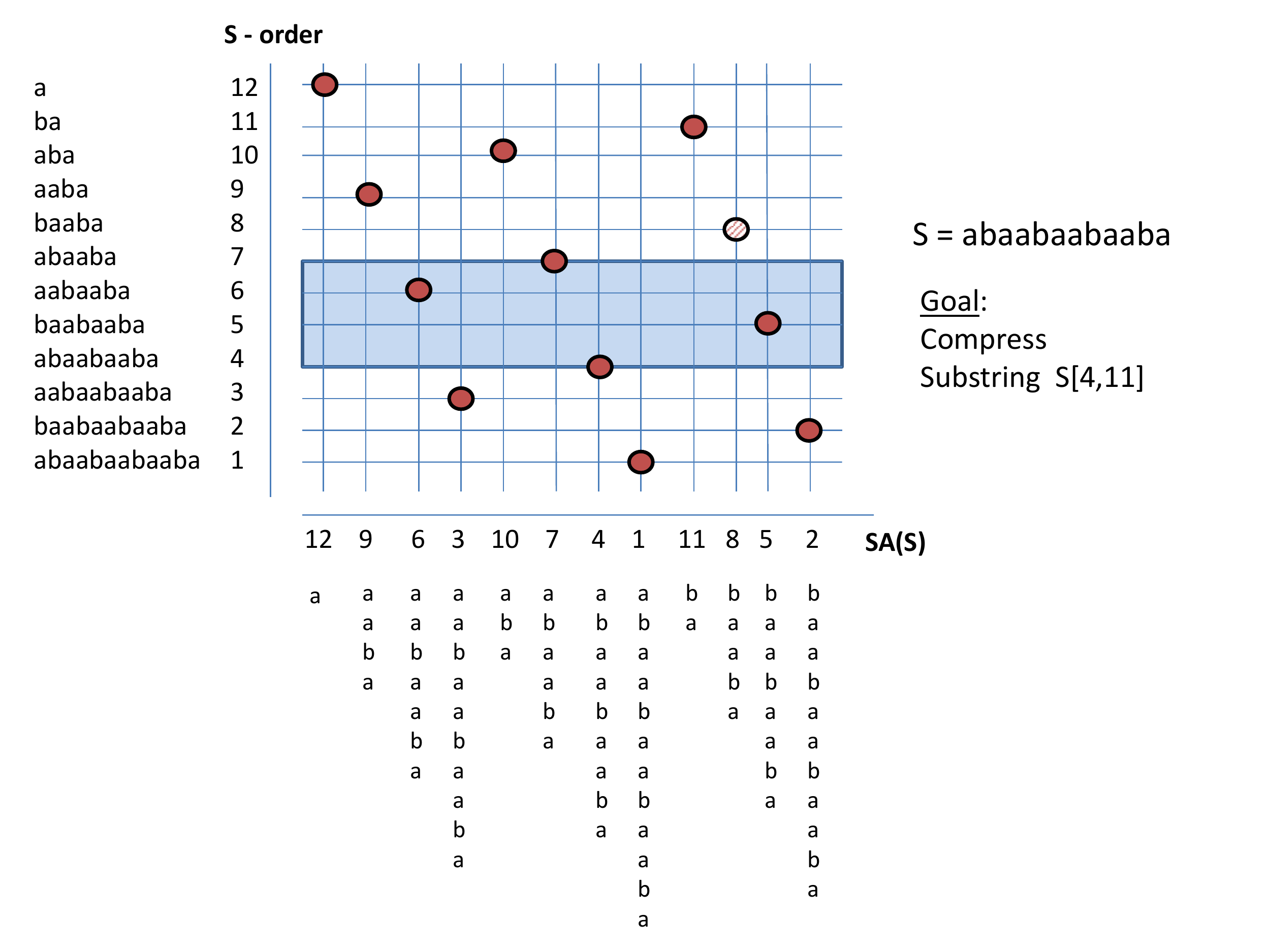}
\caption{The grid depicts $P$, the geometric representation of $S=abaabaabaaba$. The light point at $(8,10)$ represents the suffix $S_8$ (and 10-th in the suffix array), as the substring $baab$, yet to be encoded, starts at position $8$ in the string $S$. The grayed area of the grid represents the part of the substring which has already been encoded, that is $S[4,7]$. When finding the start location $t$, we will be limited to using points found in the gray area.
} \label{Figure:ILCP_GEO}
\end{figure}

\subsection{The Interval Longest Common Prefix to 3-Sided Sorted Range Reporting}
The reduction works as follows. Let $\SA(S)$ be the suffix array of our input string $S$ of length $n$.
We associate each suffix $S_i$ with its string index $i$ and with
its lexicographic index $\SA^{-1}(i)$. From these two we generate a pair $(x_i,y_i)$,
where $x_i=i$ and $y_i=\SA^{-1}(i)$.
We then preprocess the set $P=\{(x_i,y_i)\mid 1\leq i\leq n \} \subseteq [1,n]\times[1,n]$ for 3-sided sorted range reporting queries. An example of the
geometric representation of the scenario can be seen in Figure~\ref{Figure:ILCP_GEO}.

{\bf Computation of the ILCP}
Consider the suffix $S_k$ and the set of suffixes
$\Gamma=\{S_l,\ldots,S_{r}\}$. Since $\abs{\LCP(S_k,S_t)}=\max_{t'
\in [l,r]}\abs{\LCP(S_k,S_{t'})}$, $S_t$ is in fact the suffix
lexicographically closest to $S_k$, out of all the suffixes of the
set $\Gamma$.

We will first assume that we are searching for a suffix $S_{t_1}$, such that the suffix $S_{t_1}$ is \emph{lexicographically smaller} than $S_k$. The process for the case where the suffix chosen is lexicographically greater than
$S_k$ is symmetric. Therefore, once both are found all we will need to do is to choose the
best of both, i.e., the option yielding the greater
$\abs{\LCP(S_k,S_t)}$ value.

Since we have assumed w.l.o.g.\ that $S_{t_1}$ is
lexicographically smaller than $S_k$, we have actually assumed that
$y_{t_1} < y_k$, or equivalently, that $t_1$ appears to
the \emph{left} of $k$ in the suffix array. Incorporating the
lexicographical ranks of $S_k$ and $S_{t_1}$ into the expression, ${t_1}$ is
actually the value which maximizes the expression $\max\{ y_{t_1}
\mid l \leq {t_1} \leq r \mbox{ and }
y_{t_1}< y_k \}$. Notice that ${t_1}=x_{t_1}$.

Now consider the set $P=\{(x_i,y_i)\mid 1\leq i\leq n \}$. Assuming that indeed
$y_{t_1} < y_k$, we are interested in finding the maximal
value $y_{t_1}$, such that $y_{t_1}<y_k$, and $l \leq
x_{t_1} \leq r$. It immediately follows that the point
$(x_{t_1},y_{t_1}) \in P$ is the point in the range
$[l,r]\times [-\infty,y_k-1]$ having the maximal
$y$-coordinate, and therefore can be obtained efficiently by
obtaining the largest $y$-coordinate in the output of the 3-sided sorted range query.
 Once we
have found the point $(x_{t_1},y_{t_1})$, we have 
$t_1$, as $x_{t_1} =t_1$.

Equivalently, there exists $t_2$ such that $S_{t_2}$ is the suffix
lexicographically larger than $S_k$ and closest to it. In other
words, we assume $y_{t_2} > y_k$, or equivalently, that
$t_2$ appears to the \emph{right} of $k$ in the suffix
array. $t_2$ can be found using a symmetric procedure. An example of the queries performed can be seen in Figure~\ref{Figure:LCP_QUERY_GEO}.

\begin{figure}[h!]
\centering
\includegraphics[scale=0.5]{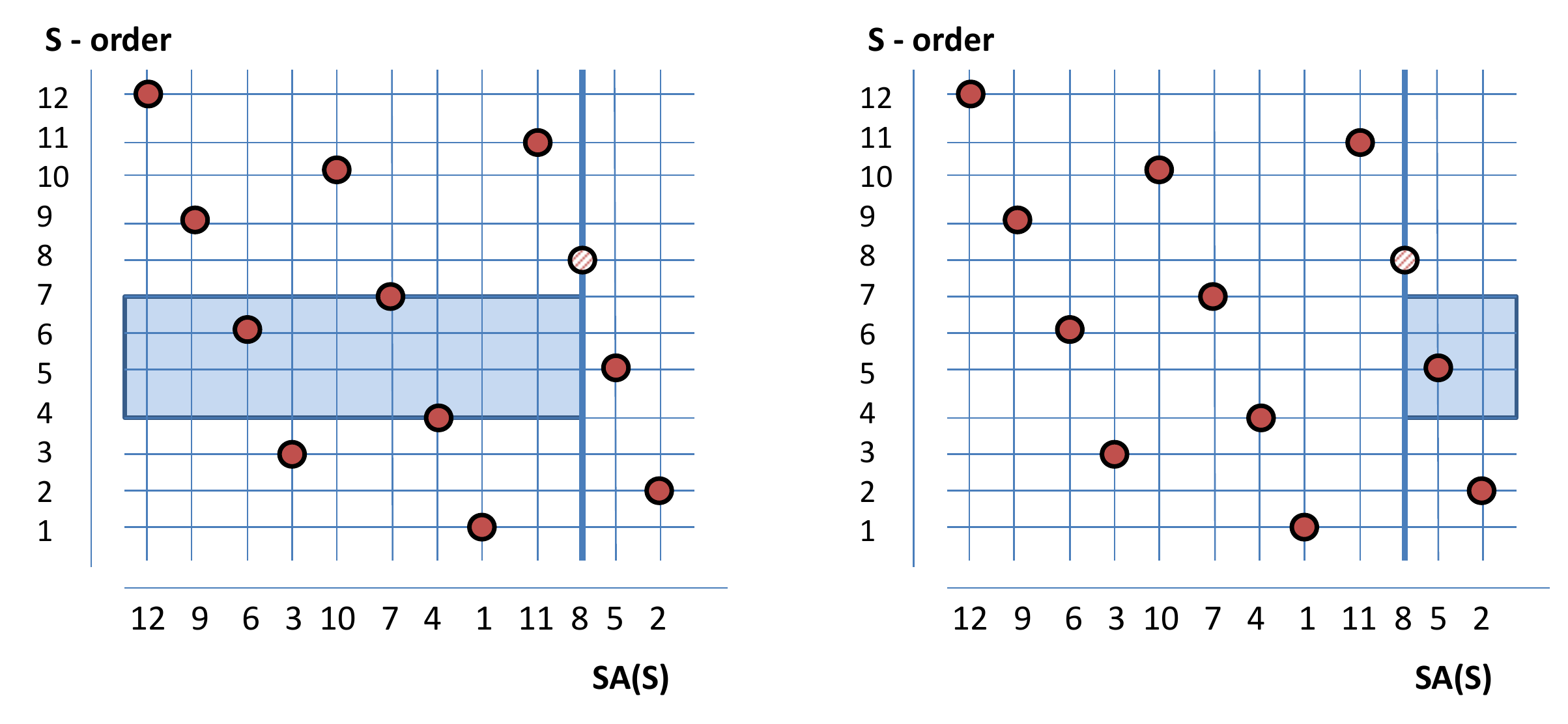}
\caption{Continued example from Figure~\ref{Figure:ILCP_GEO} with string string $S=abaabaabaaba$. The grid on the right-hand side depicts the 3-sided range successor query for $[l,r] \times [-\infty,y_k-1]$, where the grid on the left-hand side depicts the $[l,r] \times[y_k+1, \infty]$ query. In both queries the values given for the example queries are: $k=8$, $l=4$ and $r=7$. $S[l,r]$ is the substring that has already been encoded. The query $[l,r]\times [-\infty,y_k-1]$ outputs $(4,7)$ and the query on $[l,r] \times[y_k+1, \infty]$ outputs $(5,11)$. $S_5$ is chosen since $\abs{\LCP{S_8,S_5}} > \abs{\LCP{S_8,S_4}}$.
} \label{Figure:LCP_QUERY_GEO}
\end{figure}

Determining whether $t=t_1$ or $t=t_2$ is implemented by calculating both
$\abs{\LCP(S_k,S_{t_1})}$ and $\abs{\LCP(S_k,S_{t_2})}$, and
choosing the larger of the two. This gives us phrase $\rho_i$. To finish simply reiterate. Hence,

\begin{theorem}
Let $T$ be a text of length $n$. We can preprocess $T$ in $O(s(n))$ space so that we can answer substring compression queries $Q$ in $O(|\LZ(Q)|qt(n))$ time, where $s(n)$ and $qt(n)$ are the space and query times mentioned above ($occ = 1$).
\end{theorem}

\subsubsection{The GSCQ problem}

The {\em generalized substring compression} solution is more involved and uses
binary searches on suffix trees applying range queries (3-sided range successor queries and emptiness queries) during the binary
search. The interested reader should see~\cite{KKLL09}.

\subsubsection{Other applications}

The reduction from this section to range searching structures, i.e. the set of
$P=\{(x_i,y_i)\mid 1\leq i\leq n \} \subseteq [1,n]\times[1,n]$, defined by a suffix $S_i$
with $x_i=i$ and $y(i)=\SA^{-1}(i)$ had been considered beforehand. 

This reduction was first used, to the best of our knowledge, by Ferragina~\cite{F97} as
part of the scheme for searching in a dynamic text indexing scheme. The reduction and point set
were used also in {\em position restricted substring search}~\cite{Makinen2006}. However, in both {\em range reporting} was used. 

To the best of our knowledge, the first use of 3-sided range successor queries for text indexing was for {\em range non-overlapping indexing and successive list indexing}~\cite{KKL07} and in parallel for position restricted substring search in~\cite{CIR07}. 

See Section~\ref{Sec:RangeRestricted} for more range-restricted string search problems. 
\section{Top-$k$ Document Retrieval} 
\label{sec:TopK}

The {\em Top $k$ Document Retrieval} problem is an extension of the Document Retrieval problem described in Section~\ref{Sec:Document-Retrieval}. The extension is to find the {\em top $k$} documents in which a given pattern $Q$ appears, under some relevance measure. Examples of such relevance measures are (a) {\em tf(Q,d)}, the number of times $Q$ occurs in document $d$, (b) {\em mind(Q,d)}, the minimum distance between two occurrences of $Q$ and $d$, (c) {\em docrank(d)}, an arbitrary static rank assigned to document $d$. In general, the type of relevance measures which which we shall discuss here are those that are defined by a function that assigns a numeric weight $w(S,d)$ to every substring $S$ in document $d$, such that $w(S,d)$ depends only on the set of starting positions of occurrences of $S$ in $d$. We call such a relevance measure a {\em positions based relevance measure}.

The following theorem is the culmination of the work of Hon, Shah and Vitter~\cite{HSV09} and of Navarro and Nekrich~\cite{NN:TopK:12}.

\begin{theorem}
Let $D$ be a collection of strings (documents) of total length $n$,  and let $w(S,d)$ be a positions based relevance measure for the documents $d\in D$. Then there exists an $O(n)$-word space data structure that, given a string $Q$ and an integer $k$ reports $k$ documents $d$ containing $Q$ of highest relevance, i.e. with highest $w(Q,d)$ values, in decreasing order of $w(Q,d)$, in $O(|Q|+k)$ time.
\end{theorem}

Hon, Shah and Vitter~\cite{HSV09} reduced this to a problem on arrays and achieved query time of $O(|Q|+k\log k)$. We will outline their idea within this section. Navarro and Nekrich~\cite{NN:TopK:12} then showed how to adapt their solution to a 3-sided 2 dimensional range searching problem on weighted points. The solution of the range searching problem given in~\cite{NN:TopK:12} builds upon earlier work on {\em top $k$ color queries for document retrieval}~\cite{KN11}, another interesting result. We will describe the adaptation and range searching result shortly.

\subsection{Flattening the Top-$k$ Document Retrieval Suffix Tree}

Consider a generalized suffix tree $\ST$ for the document collection $D$. The leaves have a one-one correspondence with the locations within the documents. If leaf $l$ is associated with location $i$ of document $d$, we say that it is a $d$-leaf. Let $l_1,l_2,\ldots, l_q$ be the set of $d$-leaves. Then a node is a $d$-node if (a) it is a $d$-leaf or if (b) it is an internal node $v$ of $\ST$ such that it is the lowest common ancestor of adjacent $d$-leaves $l_i$ and $l_{i+1}$. Let $v$ be a $d$-node. If $u$ is is the lowest common ancestor of $v$ that is a $d$-node we say that $u$ is $v$'s $d$-parent (and that $v$ is $u$'s $d$-child). If there is no lowest common ancestor of $v$ which is a $d$-node then the $d$-parent will be a dummy node which is the parent of the root. One can easily verify that the set of $d$-nodes form a tree, called a $d$-tree, and that an internal $d$-node has at least two $d$-children. It is also straightforward to verify that the lowest common ancestor of any two $d$-nodes is a $d$-node. Hence,

\begin{lemma}~\label{lem:unique-doc}
Let $v$ be a node in the generalized suffix tree $\ST$ for the document collection $D$. For every document $d$ for which the subtree of $v$ contains a $d$-leaf there is exactly one $d$-node in the subtree of $v$ that has a $d$-parent to an ancestor of $v$.
\end{lemma}

\proof
1. Every $d$-parent of a $d$-node in the subtree of $v$ is either in the subtree or is an ancestor of $v$. 2. Assume, by contradiction, that there are two $d$-nodes $x_1$ and $x_2$ in the subtree of $v$ each with a $d$-parent that is an ancestor of $v$. However, their lowest common ancestor, which must be a $d$-node, is no higher than $v$ itself (since $v$ is a common ancestor). Hence, their $d$-parents must be in $v$'s subtree a contradiction.

Hence, since every $d$-node has a $d$-parent, there must be exactly one $d$-node with a $d$-parent to an ancestor of $v$.
\qed

\begin{corollary}~\label{cor:unique}
Let $v$ be an arbitrary node in $\ST$ and let $u$ be a descendant of $v$ such that $u$ is a $d$-node and its $d$-parent is an ancestor of $v$. Then all $d$-nodes $u'$ which are descendants of $v$ are also descendants of $u$.
\end{corollary}

\bigskip

A node in $\ST$ may be a $d$-node for different $d$'s, say for $d_{i_1}, \ldots, d_{i_r}$.  Nevertheless, since every internal $d$-node has at least two $d$-children, the $d$-tree is linear in the number of $d$-leaves and, hence, the collection of all $d$-trees is linear in the size of the $\ST$ which is $O(n)$.

In light of this in~\cite{HSV09} an array $A$ was constructed by a pre-order traversal of the tree $\ST$ such that for each node $v$ which is a $d$-node for $d\in \{d_{i_1}, \ldots, d_{i_r}\}$  indexes $j+1$ to $j+r$ are allocated in the array and contain the $d_{i_1}$-parent of $v$, $\ldots$, the $d_{i_r}$-parent of $v$. The integer interval $[l_v,r_v]$  denotes the interval bounded by the minimal and maximal indexes in $A$ assigned to $v$ or its descendants. Values $l_v$ and $r_v$ are stored in $v$.

The array $A$ was used to obtain the query result of $O(|Q|+k\log k)$ in~\cite{HSV09}. We now show how this was used in~\cite{NN:TopK:12}.

\subsection{Solving with Weighted Range Searching}

Let $j+t$ be the index in $A$ associated with $d_{i_t}$-node $v$ for $d_{i_t}\in \{d_{i_1}, \ldots, d_{i_r}\}$ and with its $d_{i_t}$-parent $u_t$. Let $S$ be the string such that the locus of $S$ is $v$. We generate a point $(j+t, depth(u_t))$, where $depth$ denotes the depth of a node in the $\ST$. The weight of the point $p$ is $w(S,d_{i_t})$. Note that all points have different $x$-coordinates and are on an integer $n \times n$ grid.

It is still necessary to store a mapping from the $x$-coordinates of points to the document numbers. A global array of size $O(n)$ is sufficient for this task.

{\bf Queries.}
To answer a top-$k$ query $Q$ first find the locus $v$ of $Q$ (in $O(|Q|)$ time). Now, by Lemma~\ref{lem:unique-doc} for each document $d$ containing $Q$ there is a unique $d$-node $u$ which is a descendant of $v$ with a $d$-parent who is an ancestor of $v$. By Corollary~\ref{cor:unique} $w(Q,d) = w(S,d)$, where $S$ is the string with locus $u$, and $w(S,d)$ is the weight of the point corresponding to the pointer from $u$ to its $d$-parent.
So, there is a unique point $(x,y)$ with $x\in [l_v,r_v]$ and $y\in[0,depth(v)-1]$ for every document $d$ that contains $Q$. Therefore, it is sufficient to report the $k$ heaviest weight nodes in $[l_v,r_v]\times [0,depth(v)-1]$. To do so Navarro and Nekrich~\cite{NN:TopK:12} proposed the {\em three sided top-$k$ range searching} problem.

\bigskip
\begin{tabular}{|l   l|}
  \hline
  \multicolumn{2}{|l|}{\bf Three sided top-$k$ range searching} \\
  \hline
{\bf \ \ Input:}& A set of $n$ weighted points on an $n\times n$ grid $G$. \\
{\bf \ \ Output:\ \ \ }& A data structure over $G$ supporting the following queries.\ \ \ \ \\
{\bf \ \ Query:}& Given $1 \leq k, h\leq n$ and $1 \leq a \leq b \leq n$ return the $k$\\ &  heaviest weighted points in the range $[a,b]\times [0,h]$.\\
\hline
\end{tabular}

\bigskip
In~\cite{NN:TopK:12} a solution was given that uses $O(n)$-word space and $O(h+k)$ query time. This result is similar to that of~\cite{KN11}. It is easy to note that for the application of top-$k$ document retrieval this yields an $O(depth(v)+k)$ query time, which is $O(|Q|+k)$, an optimal solution.

\subsection{External Memory Top-k Document Retrieval}

Lately, a new result for top-$k$ document retrieval for the external memory model has appeared in~\cite{SSTV12}. The result is I/O optimal and uses $O(n\log^* n)$ space.

More on research in the vicinity of top-k document retrieval can be found in~\cite{GN-1304-6023}. See also~\cite{BG05} to see how to add rank functionality to a suffix tree. 
\section{Range Restricted String Problems}~\label{Sec:RangeRestricted}

Research inspired by the problem of applying string problems limited to ranges has been of interest in the pattern matching community from around 2005. Some of the results are general. Others focus on specific applications. One such application  is the {\em substring compression} problem that was discussed in Section~\ref{sec:substring-compression}. These problems are natural candidates for range searching solutions and indeed many of them have been solved with these exact tools.

The first three results on range restricted variants of text indexing appeared almost in parallel. The results were for {\em property matching} (the conference version of~\cite{ACIKZ08}), {\em substring compression}~\cite{CM05} and {\em position-restricted substring searching}~\cite{Makinen2006}.

\subsubsection{Property matching} is the problem of generating a text index for a text and a collection of ranges over the text. The subsequent pattern queries asks for the locations where the text appears and are fully contained in some interval. The initial definition was motivated by {\em weighted matching}. In {\em weighted matching} a text is given with probabilities on each of the text symbols and each pattern occurrence in the text has {\em weight} which is the multiplication of the probabilities on the text symbols associated with that occurrence. Weighted matching was reduced to property matching. In~\cite{ACIKZ08} a solution was given using $O(n)$ space, where $n$ is the text size, such that queries are answered in $O(|Q|+occ_{\pi})$ time, where $Q$ is the pattern query and $occ_{\pi}$ is the number of appearances within the interval set $\pi$. The preprocessing time was near optimal and in a combination of a couple of papers was solved in optimal time~\cite{IR08,JLW09}. See also~\cite{CKWIR10}. In~\cite{Kopelowitz10} property matching was solved in the dynamic case, where intervals can be inserted and removed. Formally, $\pi$ denotes the collection of intervals and the operations are:

\begin{itemize}
\item Insert($s$, $f$) - Insert a new interval ($s$, $f$) into $\pi$.
\item Delete($s$, $f$) - Delete the interval ($s$, $f$) from $\pi$.
\end{itemize}

In~\cite{Kopelowitz10} it was shown how to maintain a data structure under interval deletions. Queries are answered in $O(Q|+occ_{\pi})$ time and deletions take $O(f-s)$ time. If both insertions and deletions are allowed then the insertion/deletion time is $O(f-s + \log\log n)$, where $n$ is the text length.

In~\cite{HPST11} a succinct version was given that uses a compressed suffix array (CSA). The solution has a multiplicative logarithmic penalty for the query and update time.

\subsubsection{Position-restricted substring searching} is the problem where the goal is to preprocess an index to allow range-restricted queries. That is the query consists of a pattern query $Q$ and a range described by text indices $i$ and $j$. This is different from property matching because the interval is not given a-priori. On the other hand, it is one interval only. The queries considered in~\cite{Makinen2006} are {\em position-restricted reporting} and {\em position-restricted counting}. Another two related queries also considered are {\em substring rank} and {\em substring select}, which are natural extensions of rank and select~\cite{BM99,GMR06,Jacobson89}. These are defined as follows.

\begin{enumerate}
\item
\textbf{PRI-Report:} Preprocess text $T=t_1\cdots t_n$ to answer queries Report$(Q=q_1\cdots q_m, i, j)$, which reports all occurrences of $Q$ in $t_i\ldots t_j$.
\item
\textbf{PRI-Count:} Preprocess text $T=t_1\cdots t_n$ to answer queries Count$(Q=q_1\cdots q_m, i, j)$, which returns the number of occurrences of $Q$ in $t_i\ldots t_j$.
\item
\textbf{Substring Rank:} Preprocess text $T=t_1\cdots t_n$ to answer queries SSR$(Q=q_1\cdots q_m, k)$, which returns the number of occurrences of $Q$ in $t_1\cdots t_k$.
\item
\textbf{Substring Select:} Preprocess text $T=t_1\cdots t_n$ to answer queries SSS$(Q,k)$, which returns the $k^{th}$ occurrence of $Q$ in $T$.
\end{enumerate}

We note that substring rank and position-restricted counting reduce to each other. Also, position-restricted reporting can be obtained from applying one substring-rank and $occ_{i,j}$ substring-selects, where $occ_{i,j}$ is the number of pattern occurrences in $t_i\ldots t_j$. We leave it to the reader to verify the details.



\bigskip

\noindent
\uline{Reporting:} For the reporting problem M{\"a}kinen and Navarro~\cite{Makinen2006} reduced the problem to range reporting. The way to do so is to first find the $SA$-range of $Q$. Then this range and the position-restricted range $i$ to $j$ define a rectangle for which range reporting is used. One can use any of the data structures mentioned in Section~\ref{sec:oneError}. For example with $O(n\log^{\epsilon}n)$ space, for any constant $\epsilon$, one can answer queries in $O(m + \log\log n + occ_{i,j})$ (we assume the alphabet is from $[1,n]$ otherwise if it is from $[1,U]$ then there is an extra additive factor of $\log\log U$). Crochemore et al.~\cite{CIR07} noticed that a different type of reduction, namely range next value, could be more useful. The authors of~\cite{Makinen2006} were more concerned with space issues. So, they also proposed a data structure which uses n + $o(n)$ space and reports in $O(m+\log n + occ_{i,j})$.  The reporting time was improved by Bose et al.~\cite{BHMM09} to $O(m+\log n / \log\log n + occ_{i,j})$ with the same space constraints. Yu et al.~\cite{YHW11} suggested a different algorithm with the same space and reporting time, but were able to report the occurrences in their original order.

Bille and G{\o}rtz~\cite{BG11} went on to show that with $O(n(\log^{\epsilon}n+\log\log U))$ space the query time can be improved to $O(m + occ_{i,j})$, which is optimal. They also solved position-restricted reporting merged with property matching. The space and query time remain the same.

An interesting result for the reporting variant appeared in~\cite{HSTV12}. Specifically, it was shown that a succinct space $\polylog n$ time index for position-restricted substring searching is at least as hard as designing a linear space data structure for 3D range reporting in $\polylog n$ time.

\bigskip

\noindent
\uline{Counting:} In~\cite{Makinen2006} the same data structure that uses n + $o(n)$ space and reports in $O(m+\log n)$ was used for counting\footnote{There is another result there that assumes faster query times that is flawed. See the introduction in~\cite{KLP11} for an explanation.}. Once again, Bose et al.~\cite{BHMM09} can improve the counting time to $O(m+\log n / \log\log n)$. Kopelowitz et al.~\cite{KLP11} presented a counting data structure that uses $O(n(\log n/ \log\log n))$ space and answers counting queries in time $O(m + \log\log|\Sigma|)$. Recently, in the upcoming journal version of~\cite{BG11} a similar result appears. The space is the same. However, the counting time is $O(m + \log\log n)$, which can be slightly worse. Lately, Gagie and Gawrychowski~\cite{CORR-GG-12} showed that if the alphabet is of size $\polylog (n)$ then $O(n)$ space can be achieved. Moreover, if the alphabet size $\sigma = \log^{O(1)} n$ then they can reduce the query time to an optimal $O(m)$.

\bigskip

\noindent
\uline{Substring Select:} In~\cite{Makinen2006} a solution for indexing for substring select is given. The space is $O(nK\log{\sigma}/\log n)$, where $K$ is an upper bound on the size of the queried patterns. The query time is $O(m\log{\sigma}/\log\log n)$. This was improved in~\cite{KLP11} to allow for any length query with  $O(n\log n/\log\log n)$ space and optimal $O(m)$ query time. The proposed solution in~\cite{KLP11} uses persistent data structure which is a basic ingredient in most of the range searching solutions.

\subsubsection{Substring compression} has been expanded on in Section~\ref{sec:substring-compression}. It was introduced in~\cite{CM05} and improved upon in~\cite{KKLL09}. The results are detailed in Section~\ref{sec:substring-compression}. One of the problems that is of interest in substring compression is the $ILCP$ (interval longest common prefix) query. This inspired Amir et al.~\cite{AALLLP11} to consider extensions to LCP range queries of different types.

\subsubsection{Range non-overlapping indexing and successive list indexing~\cite{KKL07}}. In range non-overlapping indexing one wants to prepare an index so that when give a pattern query one can return a maximal set of occurrences so that the occurrences do not overlap. In successive list indexing one prepares an index to answer queries where a pattern is given along with a position $i$ and one desires to find the first occurrence of the pattern after $i$. A reduction to range successor was used to solve this problem. Along with the results of range sorted reporting~\cite{NN12} one can solve the former with space $O(n\log^{\epsilon}n)$ space and $O(\log\log n + occ)$ query time. For the latter the query time is $O(\log\log n)$.

\subsubsection{Range Successor in 2D} solves several of the problems mentioned in this section. This has been discussed in Section~\ref{sec:substring-compression} and is referred to in~\cite{CIKRTW12,NN12,YHW11}. It is interesting that its generalization sorted range reporting~\cite{NN12} is a variant of range reporting that was considered in the community because of the unique range search problems that arise.

\section{Lower Bounds on Text Indexing via Range Reporting}
\label{sec:LowerBound}

A novel use of range searching is its use to show lower bounds on text indexing via reductions from range reporting~\cite{CHSV08}.

\begin{theorem}
Let $S=\{(x_1,y_1), (x_2,y_2), \ldots, (x_n,y_n)\}$ be a set of $n$ points in $[1,n]\times [1,n]$. We can construct a text $T$ of length $O(n\log n)$ bits along with $O(n\log n)$ bits of auxiliary data such that we can answer range reporting queries on $S$ with $O(\log^2n)$ pattern match queries on $T$, each query is a pattern of length $O(\log n)$.
\end{theorem}

We denote the set of $x$-coordinates, of $S$, $X = \{x_1,\ldots, x_n\}$ and the set of $y$-coordinates, of $S$, $Y = \{y_1,\ldots, y_n\}$.

The idea is as follows. Each point $(x,y)\in [1,n]\times [1,n]$. So, $x$ and $y$ both have binary representations of $\log n$ bits\footnote{We assume that $n$ is a power of 2. Otherwise, it will be $\lfloor \log n \rfloor + 1$.}. Denote with $b(w)$ the binary representation of a number $w$ and the reverse of the binary representation with $b^R(w)$. The text $T$ (from the theorem) constructed is $b^R(x_1)\#b(y_1)\$b^R(x_2)\#b(y_2)\$\ldots\$b^R(x_1)\#b(y_1)$.

To obtain the result a collection of pattern queries on $T$ is generated whose answers will yield an answer to the range reporting problem on the point set $S$. To this end, sort $b(y_1),\ldots, b(y_n)$ and $b^R(x_1),\ldots, b^R(x_n)$. Let $\pi$ denote the ordering of the former and $\tau$ denote the ordering of the latter, i.e. $b(\pi(y_1)) < \ldots < b(\pi(y_n))$, where $<$ is a lexicographic-less than, and $b^R(\tau(x_1)) < \ldots < b^R(\tau(x_n))$. The $n$-length arrays $A = <b^R(\tau(x_1)), \ldots, b^R(\tau(x_n))>$ and $B = <b(\pi(y_1)), \ldots, b(\pi(y_n))>$ will be the basis of the search.

Over each of the arrays construct a binary search tree with each node representing a range of elements. Without loss of generality, consider the binary tree over $B$. The root represents all elements of $Y$. The left son is associated with one bit 0 and represents $R_0 = \{y \in Y | 0$ is prefix of $b(y)\}$ and the right son represents $R_1 = \{y \in Y | 1$ is prefix of $b(y)\}$ - each is a range over $B$ - check. The left son of the left son of the root represents $R_{00} = \{y \in Y | 00$ is prefix of $b(y)\}$, etc. In general, each node is associated with a binary string, say $b_0\ldots b_i$, formed by the walk down from the root to the node and is also associated with a range, which we call a {\em node-range},  $R_{b_0\ldots b_i} = \{y \in Y | b_0\ldots b_i$ is prefix of $b(y)\}$. The number of nodes in the binary tree and, hence, the number of ranges is $\leq 2n-1$. Each range can be represented as a pair of indexes to the array. Hence, the size of the auxiliary information is $O(n)$-words, or $O(n\log n)$ bits. We construct a complementary binary tree for $A$, with ranges $RX$.

An easy well known observation is that any range $R^{q,r} = \{y | 1 \leq q \leq y \leq r \leq n \}$  can be expressed as the disjoint union of at most $2\log n$ node-ranges. The node-ranges of the disjoint union can be found by a traversal up and down the binary tree using the binary representations of $q$ and $r$.

Now consider a range query on $S$, say $[x_{left}, x_{right}]\times [y_{bottom},y_{top}]$. This can be seen as a query for all $(x,y)$ such that $y \in R^{y_{bottom},y_{top}}$ and $x \in RX^{x_{left},x_{right}}$. By the previous observation this can be transformed into $O(\log^2n)$ queries for all $(x,y)$  such that $x$ is in one of the node-ranges in the disjoint union expressing $R^{y_{bottom},y_{top}}$ and $y$ is in one of the node-ranges in the disjoint union expressing $RX^{x_{left},x_{right}}$.

We show an a indexing query that searches for all $(x,y)$ such that $x \in RX_{c}$ and $y \in R_d$ both node-ranges, the former for the binary string $c$ over the array $A$ and the latter for the binary string $d$ over the array $B$. We define a pattern query $Q = c^R\#d$. We query the text index with $Q$. Every location where $Q$ appears corresponds to a point that is in the desired range as $c^R$ must align with the end of an $b^R(x_i)$, which is the same as $c$ being a prefix of $b(x_i)$ and $d$ is a prefix of $(y_i)$, which is exactly the desired.

Chazelle~\cite{Cha90a} showed that in the pointer machine model an index supporting 2D range reporting in $O(\polylog(n) + occ)$ query time, where $occ$ is the number of occurrences, requires $\Omega(n(\log n/\log\log n))$ words of storage. Hence,

\begin{theorem}
In the pointer machine model a text index on $T$ of size $n$ which returns locations of pattern occurrences in $O(\polylog(n)+ occ)$ time requires $\Omega(n(\log n/\log\log n))$ bits.
\end{theorem}

\subsubsection{More on this result and related work}
 In~\cite{CHSV08} there are also very interesting results reducing text indexing to range searching. The reduction is known as a Geometric BWT, transforming a BWT into a point representation.  The reductions in both directions show that obtaining improvements in space complexity of either will imply space complexity improvements on the other.

Another two lower bounds, along the lines of the lower bound described here, are for {\em position restricted substring search}~\cite{HSTV12}, for {\em forbidden patterns}~\cite{FGKLMSV12} and for {\em aligned pattern matching}~\cite{Thankachan11}.

\subsubsection{Appreciation.}

I wanted to thank my numerous colleagues who were kind enough to provide insightful comments on an earlier version and pointers to work that I was unaware of. These people include (in alphabetical order) Phillip Bille, Timothy Chan, Francisco Claude, Pooya Davoodi, Johannes Fischer, Travis Gagie, Roberto Grossi, Orgad Keller, Tsvi Kopelowitz, Muthu Muthukrishnan, Gonzalo Navarro, Yakov Nekrich, Rahul Shah, Sharma Thankachan,  Rajeev Raman, and Oren Weimann. Special thanks to Orgad, Rahul, Sharma and Yakov for numerous Skype conversations in which I learned more than can be contained within this monologue.




\bibliographystyle{abbrv}
\bibliography{ref}

\end{document}